\documentclass[12pt,letterpaper]{article}
\usepackage{CJKutf8}
\pdfoutput=1
\usepackage{booktabs}
\usepackage{siunitx}
\usepackage{tikz}
\usepackage{geometry}
\usepackage{tikz}
\usepackage{color}
\usepackage{tcolorbox}
\usepackage{slashed}
\definecolor{refkey}{rgb}{1,0,0}
\definecolor{labelkey}{rgb}{0,0,1}
\usepackage{amsmath}
\usepackage{bbold,amsfonts,amssymb,amsmath}
\usepackage{esint}
\usepackage{rotating}
\usepackage[utf8]{inputenc}
\usepackage{cite}
\usepackage{graphicx}
\usepackage{caption}
\usepackage{subcaption}
\usepackage{booktabs}

\usepackage{listings}

\usepackage{hyperref}

\usepackage{soul}
\setstcolor{red}

\usepackage{tikz}
\usetikzlibrary{calc}

\newcommand{\be}{\begin{equation}}
	\newcommand{\ee}{\end{equation}}
\newcommand{\ben}{\begin{displaymath}}
	\newcommand{\een}{\end{displaymath}}
\newcommand{\bea}{\begin{equation}\begin{aligned}}
		\newcommand{\eea}{\end{aligned}\end{equation}}
\newcommand{\bean}{\begin{eqnarray*}}
	\newcommand{\eean}{\end{eqnarray*}}

\newcommand{\eg}{{\it e.g.}}

\newcommand{\commentout}[1]{}

\newcommand{\beq}{\begin{equation}}
	\newcommand{\eeq}{\end{equation}}
\newcommand{\beqr}{\begin{displaymath}}
	\newcommand{\eeqr}{\end{displaymath}}
\newcommand{\beqa}{\begin{eqnarray}}
	\newcommand{\eeqa}{\end{eqnarray}}
\newcommand{\beqar}{\begin{eqnarray*}}
	\newcommand{\eeqar}{\end{eqnarray*}}

\newcommand{\cF}{{\cal F}}

\newcommand{\half}{\ensuremath{\frac{1}{2}}}
\newcommand{\mpi}{\ensuremath{m_{\pi}}}
\newcommand{\fpi}{\ensuremath{f_{\pi}}}

\renewcommand{\Re}{\ensuremath{\mathrm{Re}}}

\newcommand{\MeV}{\ensuremath{\mbox{MeV}}}
\newcommand{\GeV}{\ensuremath{\mbox{GeV}}}

\newcommand{\chiPT}{\ensuremath{\chi_{PT}}}

\begin{document}
%%%%%%%%%%%%%%%%%%%%%%%%%%%%%%%%%%%%%%%%%%%%%%%%%%%%%%%%%%%%%%%%%%%%%%%%
%%%%%%%%%%%%%%%%%%%%%%%%%%%%%%%%%%%%%%%%%%%%%%%%%%%%%%%%%%%%%%%%%%%%%%%%
%%%%%%%%%%%%%%%%%%%%%% TITLEPAGE %%%%%%%%%%%%%%%%%%%%%%%%%%%%%%%%%%%%%%%
%%%%%%%%%%%%%%%%%%%%%%%%%%%%%%%%%%%%%%%%%%%%%%%%%%%%%%%%%%%%%%%%%%%%%%%%
%%%%%%%%%%%%%%%%%%%%%%%%%%%%%%%%%%%%%%%%%%%%%%%%%%%%%%%%%%%%%%%%%%%%%%%%

\title{\Large \bf Gauge Theory Bootstrap with varying $N_c$}

\author{
	Rafael Cordoba$^\text{1}$, Yifei He$^\text{1}$, \\
	Martin Kruczenski$^\text{2}$, Derek Ping$^\text{2}$, Jignesh Mohanty$^\text{2}$, Emil Alegria$^\text{2}$  \thanks{\texttt{ rafael.cordoba@ens.fr, yifei.he@ens.fr,\\ markru@purdue.edu, jmohanty@purdue.edu, dping@purdue.edu, ealegria@purdue.edu}} \\
[2.0mm]
$^1$ \small Laboratoire de Physique de l'\'Ecole Normale Supérieure, ENS, Université PSL,\\
\small CNRS, Sorbonne Université, Université Paris Cité, F-75005 Paris, France \\
$^2$ \small Department of Physics and Astronomy, Purdue University, West Lafayette, IN 47907, USA.}

\date{}

\maketitle

\begin{abstract}
We use the Gauge Theory Bootstrap to study asymptotically free gauge theories with a varying number of colors $N_c$. The parameter $N_c$ enters the computation through the sum rules and asymptotic form factors, allowing us to tune this microscopic parameter directly and examine the corresponding low-energy strongly coupled dynamics. For $3\le N_c\le 12$, we extract the scattering lengths, effective ranges, and low-energy constants $\bar{l}_i$, and study their $N_c$ dependence. We also obtain the $\sigma$, $\rho$, and $f_2$ resonances for the same range of $N_c$. The $\rho$ and $f_2$ masses depend only weakly on $N_c$, and their widths decrease as $N_c$ increases, consistent with the $1/N_c$ scaling expected for conventional $q\bar{q}$ mesons. In contrast, the $\sigma$ meson mass and width both increase with $N_c$, in agreement with a previous unitarized $\chi$PT analysis. These results provide a further test of the Gauge Theory Bootstrap for studying asymptotically free gauge theories.
\end{abstract}

\clearpage

\tableofcontents

\newpage

\section{Introduction}

Determining the long distance physics from the microscopic parameters of a physical system is an important problem in theoretical physics. One of the most prominent examples is the theory of strong interactions. In this case, the UV theory is given by an asymptotically free gauge theory with gauge group $SU(N_c)$ and $N_f$ massive quarks, and the theory undergoes chiral symmetry breaking and confinement. The IR physics is dominated by the pseudo-Goldstone bosons -- pions -- and it is challenging to study their strongly coupled dynamics with a given set of UV parameters.

Recently, the Gauge Theory Bootstrap (GTB) method \cite{PhysRevLett.133.191601,PhysRevD.110.096001,He:2024nwd,He:2025gws} was proposed to solve this problem. The method combines modern S-matrix/form factor bootstrap numerical techniques \cite{Paulos:2016but,Paulos:2016fap,Paulos:2017fhb,Karateev:2019ymz} encoding generic consistency constraints, with the UV dynamical information from perturbative QCD through the SVZ sum rules \cite{Shifman:1978bw,SHIFMAN1979385,SHIFMAN1979448} and asymptotic form factors à la Brodsky-Lepage \cite{osti_1447331,Lepage:1980fj}; furthermore, IR constraints from weakly coupled pions \cite{PhysRevLett.17.616,GASSER1984142} are used to control the very low energy region. These ingredients, when used together, allowed the computation of the strongly coupled resonances (masses, widths) in the intermediate energy region where both weakly coupled pions and quark/gluon descriptions fail. In addition, low energy constants of the chiral effective field theory \cite{GASSER1984142} can be obtained as a result. The initial implementation of the method was carried out for $N_f=2,N_c=3$ \cite{PhysRevLett.133.191601,PhysRevD.110.096001,He:2024nwd,He:2025gws} and results compared reasonably well with experimental data, providing a proof of concept of the GTB idea. Further implementations include efficient discretizations \cite{Cordoba:2025wnv} and phenomenologically motivated analysis \cite{Zahed:2024haa}. To further establish the GTB framework, a natural next step is to study families of gauge theories with varying UV parameters and examine how their strong dynamics change. Such results can serve as a benchmark for future lattice and quantum simulations of these theories.   

In this work, we take a step in that direction and carry out the GTB computation for two quark flavors at physical pion mass, with varying number of colors $N_c$ in the range $3\le N_c\le 12$. The results summarized in section \ref{summary_low_energy} for the low energy parameters and in eqs. \eqref{summary_resonances}-\eqref{summary_resonances_last} for the resonances match well with the $N_c$ scaling predictions of the large-$N_c$ expansion of QCD \cite{tHooft:1973alw,Witten:1979kh} that provides the expected large $N_c$ scaling of the various low energy parameters, resonances, etc.  While significant progress has been made with lattice simulations \cite{DeGrand:2016pur,Hernandez:2019qed,Baeza-Ballesteros:2022azb,DeGrand:2023hzz,DeGrand:2024lvp,Baeza-Ballesteros:2025iee,bonanno2025nonperturbativedeterminationmesonmasses}, GTB serves as an alternative, efficient way to do this analysis. This, in particular, includes studying the evolution of resonance properties with $N_c$ to understand the nature of those states. Such study was previously done using indirect methods such as unitarized chiral perturbation theory \cite{Pelaez:2006nj}, giving important insights into, e.g. the nature of the $\sigma$ meson. In the GTB approach, the number of colors of the UV theory is an input parameter thus putting the resonance properties and the IR predictions in direct relation with the microscopic physics. The basic setup of the $N_c$ computation is the same as the one in the initial implementation \cite{PhysRevLett.133.191601,PhysRevD.110.096001,He:2024nwd,He:2025gws}. In this setup, the number of colors enters explicitly as a free UV input parameter in the QCD sum rules and the asymptotic form factor. Meanwhile, $N_c$ also enters implicitly through the pion decay constant $f_\pi$ (defined through axial current matrix element) which, at leading order in $N_c$, behaves as $f_\pi\sim \sqrt{N_c}$. Finally, the results obtained here for $N_c\le 12$ can be extrapolated to large $N_c$ limit where the theory simplifies and mesons become stable. The latter also connects with recent work \cite{Albert:2026xyz} where the GTB idea of combining bootstrap constraints with UV information was implemented in the strict $N_c\rightarrow \infty$ limit.

\smallskip

The paper is organized as follows. In section \ref{GTBreview}, we briefly review the GTB setup. In the follow section, we specify how the $N_c$ parameter is varied in GTB. Sections \ref{LEC} and \ref{res} show the results for the low energy parameters and the resonances as functions of $N_c$. We conclude in the final section and give supplementary materials, including useful formulas, and details on the numerical implementations in the appendices.
As in \cite{He:2024nwd,He:2025gws}, the code for the GTB program with varying $N_c$ is publicly available on GitHub at \href{https://github.com/hyfysics/gauge-theory-bootstrap}{gauge-theory-bootstrap}.

\section{Brief review of Gauge Theory Bootstrap}\label{GTBreview}

We start by considering 2-to-2 pion scattering:
\begin{equation}
\pi_a+\pi_b\to \pi_c+\pi_d
\end{equation}
with an amplitude written as function of the Mandelstam variables $s$, $t$, $u$, ($s+t+u = 4 m_\pi^2$) and given by
\begin{equation} \begin{array}{ccl}
    T_{ab,cd} &= &A(s,t,u) \delta_{ab} \delta_{cd}+ A(t,s,u) \delta_{ac}\delta_{bd} + A(u,t,s) \delta_{ad}\delta_{bc}\\
    &&\\
    &=& {1 \over 3}T^{I=0} \delta_{ab} \delta_{cd} + {1 \over 2} T^{I=1} ( \delta_{ac}\delta_{bd} - \delta_{ad} \delta_{bc}) +{1 \over 2} T^{I=2}(\delta_{ac}\delta_{bd}+\delta_{ad}\delta_{bc} -{2 \over 3} \delta_{ab}\delta_{cd})
    \end{array}\label{amp}
\end{equation}
Furthermore, $A(s,t,u)$ satisfies crossing $A(s,t,u)=A(s,u,t)$ and Mandelstam analyticity, namely it is an analytic function of $(s,t,u)$  whose only singularities are cuts for $s,t,u\ge 4m_\pi^2$.  The fixed-isospin amplitudes $T^{I=0,1,2}$ are related to the partial waves through:
\begin{equation}\label{pw}
    f_\ell^I(s) = {1 \over 4} \int_{-1}^1 d \mu P_\ell(\mu)T^I(s,t),\hspace{0.75cm}t=\frac{(4m_\pi^2-s)(1-\mu)}{2}
\end{equation}
where the partial amplitudes for physical energies are
\begin{equation}
    S^I_\ell(s) = 1 + i \pi \sqrt{s-4 m_\pi^2 \over s} f^I_\ell(s),\;\; s\in \mathbb{R}_{\ge 4 m_{\pi}^2}
\end{equation}
and satisfy the unitarity constraints:
\begin{equation}\label{unit}
    |S^I_\ell(s)|\le 1,\;\; s>4m_\pi^2
\end{equation}
The amplitude can be parameterized in such a way that analyticity and crossing are manifest:
\begin{equation} \begin{array}{ccl}
    A(s,t,u)  &=& T_0 +{1 \over \pi} \int_{4m_\pi^2}^\infty dx\, {\sigma_1(x) \over x-s} + {1 \over \pi}  \int_{4m_\pi^2}^\infty dx \, \sigma_2(x) \left[ {1\over x-t} + {1 \over x-u} \right]\\
    &&\\ &&+ {1 \over \pi^2} \int_{4 m_\pi^2}^\infty dx \int_{4 m_\pi^2}^\infty  dy \left[{\rho_1(x,y) \over (x-s)(y-t)} + {\rho_1(x,y) \over (x-s)(y-u)} + {\rho_2(x,y) \over (x-t)(y-u)} \right] .
    \end{array}\label{amplitude}
\end{equation}
The variables $\{T_0, \sigma_{1,2},\rho_{1,2}\}$ are further subject to unitarity constraints \eqref{unit}. In the modern S-matrix bootstrap program \cite{Paulos:2016but,Paulos:2016fap,Paulos:2017fhb}, this way of reducing the amplitude to a set of variables subject to the generic constraints of analyticity, crossing and unitarity has been used systematically to put bounds on low energy parameters and explore the space of generic scattering amplitudes. In particular, this includes applications to pion physics to bound low energy constants from generic physical principles \cite{Guerrieri:2018uew,Bose:2020cod,Bose:2020shm,Guerrieri:2020bto,Albert:2022oes,Fernandez_2023,Albert:2023jtd,Albert:2023seb,Ma:2023vgc,Bocchia:2026kew,Bocchia:2026lty}.\footnote{A different direction is to use the bootstrap parameterization to fit experimental data of pion scattering \cite{Guerrieri:2024jkn}.}

Connection to UV theory is possible by coupling the partial amplitudes to the form factors and spectral densities through a positive semidefinite matrix for each set of quantum numbers and physical energies \cite{Karateev:2019ymz}:\footnote{See \cite{Karateev:2020axc,Chen:2021pgx,Correia:2022dyp,Cordova:2023wjp} for previous work on the form factor bootstrap for 2d theories.}
\beq\label{Bmt}
   	\left(
   	\begin{array}{c c c}
   		1 &\ S^I_\ell(s)\ &\ \ \cF^I_\ell(s) \\
   		S_\ell^{I*}(s) &\ 1\ & \ \ \cF^{I*}_\ell(s) \\
   		\cF^{I*}_\ell(s) &\ \cF^I_\ell(s)\ &\ \ \rho^I_\ell(s)
   	\end{array}
   	\right) \succeq 0,\;\; \forall I,\ell, \ s>4m^2_{\pi},\;\; 
\eeq
The rescaled form factor $\mathcal{F}^I_\ell(s)$ is defined as 
the overlap of the state created by a local operator with the 2-pion states of fixed total momentum $P$, isospin $I$ and angular momentum $\ell$:
\begin{equation}
{}_{\text{out}}\langle P,I,\ell|\mathcal{O}^I_\ell(x=0)|0\rangle=\mathcal{F}^I_\ell(s),\;\; s=P^2
\end{equation}
It is related to the usual 2-pion form factor $F$ by a kinematic phase space factor (see eq. \eqref{FcurlyF} below) where the latter is  defined as
\begin{equation}
{}_{\text{out}}\langle\pi^-(p_1)\pi^+(p_2)|\mathcal{O}^I_\ell(x=0)|0\rangle=F^I_\ell(s),\;\; s=(p_1+p_2)^2
\end{equation}
The form factor $F^I_\ell(s)$ is an analytic function of $s$ with a cut $s>4m_\pi^2$ and can thus be parameterized by its imaginary part on the cut:
\begin{equation}
    F_\ell^I(s) = F^I_\ell(0) + {s \over \pi} \int_{4 m_\pi^2}^\infty dx {\text{Im} F^I_\ell(x) \over x(x-s)}.
\end{equation}

The spectral density associated to a given local operator is defined as the norm of the state created by the  local operator acting on the vacuum:
\begin{equation}
    \langle0|\mathcal{O}^{I\dagger}_\ell(P)\mathcal{O}^{I}_\ell(P')|0\rangle=\rho^I_\ell(s)\delta^{(4)}(P-P'),\;\; s=P^2
\end{equation}
It is related to the time-ordered 2-point function
\begin{equation}
    \Pi^I_\ell(s)=i\int\frac{d^4 x}{(2\pi)^4}e^{ip\cdot x}\langle 0|\hat{T}\{\mathcal{O}^{I\dagger}_\ell(x)\mathcal{O}^{I}_\ell(0)\}|0\rangle
\end{equation}
through a dispersion relation such that
\begin{equation}
    2\text{Im}\Pi^I_\ell(s+i\epsilon)=\rho^I_\ell(s),\;\; s>4m_\pi^2
\end{equation}

\smallskip

Up to this point, only generic consistency constraints have been used on the amplitudes, form factors and spectral density. To specify the dynamical information from the UV gauge theory, one identifies the operator $\mathcal{O}^I_\ell$ as a QCD local current. For example for the scalar and vector ones we can consider \footnote{$\Delta_\mu=\frac{1}{\sqrt{2}}(0,1,i,0)$}
\begin{subequations}\label{currents}
	\beqa
	j_S &\equiv& m_q(\bar{u}u+\bar{d}d) \label{jscalar}\\
	j_V &\equiv& \frac{1}{2}(\bar{u}\gamma^\mu u-\bar{d}\gamma^\mu d)\Delta_\mu
    \end{eqnarray}
\end{subequations}
In the high energy region, the 2-point functions of these operators are given by perturbative QCD expressions which, to leading order, take the form:
\begin{subequations}
\begin{eqnarray} 
   		\Pi^0_0(s) &=& -\frac{s}{(2\pi)^4}\frac{ m_q^2}{8\pi^2}  \Bigg\{ 2N_c \ln\left(-\frac{s}{\mu^2}\right)+O(\alpha_s)\Bigg\}  \label{PIS}\\
		\Pi^1_1(s) &=&  -\frac{s}{(2\pi)^4}\frac{ N_c }{24\pi^2} \Bigg\{\ln\left(-\frac{s}{\mu^2}\right)+O(\alpha_s) \Bigg\} \label{PIV} 
\end{eqnarray}
\end{subequations}
See \eqref{2pt} for beyond-leading-order expressions. The perturbative QCD computations then give the finite energy sum rules \cite{PhysRevD.110.096001}
\begin{equation}
    \int_{4m_\pi^2}^{s_0} \rho(x) x^n dx = -s_0^{n+1} \int_{0}^{2\pi} e^{i(n+1)\varphi} \Pi(s_0 e^{i\varphi}) d\varphi \label{fesr}
\end{equation}
Here $s_0$ is an energy scale beyond which perturbative QCD results apply to a reasonably good approximation. In this work, $s_0$ is taken to be
\begin{equation}
    s_0\sim (2\text{ GeV})^2,\;\; \alpha_s(s_0)\sim 0.3
\end{equation}
Now, the RHS of \eqref{fesr} containing the UV information can be evaluated and provides a constraint for the spectral density integrated in the strongly coupled region $4\mpi^2\le s\le s_0$ (LHS of \eqref{fesr}). 

Additionally, near threshold, free 2-pion kinematics determines the spectral density for scalar, vector, and spin-2 currents \cite{GASSER198765} behavior to be:\footnote{As can be computed from the 2-pion terms of the currents \eqref{currents} and energy momentum tensor.}
\begin{subequations}\label{lowerho}
\begin{eqnarray}
        \rho^0_0 &\simeq& {m_\pi^4 \over (2 \pi)^4} {3 \over 16 \pi} \left( 1 - {4 m_\pi^2 \over s} \right)^{1/2}\label{rhoS0fac}\\
    \rho^1_1 &\simeq& {1 \over (2 \pi)^4} {s \over 24 \pi} \left( 1 - {4 m_\pi^2 \over s} \right)^{3/2}\\
    \rho^0_2 &\simeq& {1 \over (2 \pi)^4} {s^2 \over 160 \pi} \left( 1 - {4 m_\pi^2 \over s} \right)^{5/2}
\end{eqnarray}
\end{subequations}
There is also a similar relation between the form factor $F^I_\ell$ and the rescaled form factor $\mathcal{F}^I_\ell$:
\begin{subequations}\label{FcurlyF}
    \begin{eqnarray}
        \mathcal{F}_0^0(s) &= \frac{\sqrt{6\pi}}{16\pi^3} \frac{1}{s^{\frac{1}{4}}} \left(\frac{s-4m_\pi^2}{4}\right)^{\frac{1}{4}} F_0^0(s) \\
    \mathcal{F}_1^1(s) &= \sqrt{\frac{4\pi}{3}} \frac{1}{8\pi^3} \frac{1}{s^{\frac{1}{4}}} \left(\frac{s-4m_\pi^2}{4}\right)^{\frac{3}{4}} F_1^1(s) \\
    {\cal F}^0_2(s) &= \frac{1}{4\pi^3} \sqrt{\frac{\pi}{5}}\frac{1}{s^{\frac{1}{4}}}\left(\frac{s-4m_\pi^2}{4}\right)^{\frac{5}{4}}F^{0}_2(s)
\end{eqnarray}
\end{subequations}

At low energy or zero momentum transfer, the form factors of the conserved currents have their normalization fixed  by charge conservation. For example, for the scalar and vector currents in \eqref{currents}, the form factors satisfy
\begin{equation}\label{f0}
F^0_0(0)=m_\pi^2,\;\; F^1_1(0)=1
\end{equation}
In the UV, at asymptotically high energies, the form factors are dominated by the Brodsky-Lepage behavior \cite{osti_1447331,Lepage:1980fj} which, for the vector form factor at $N_c=3$, takes the form:
\begin{equation}\label{BL}
  F_{1}^1(s)\overset{s\to \infty}{\simeq}-\frac{16\pi\alpha_sf_\pi^2}{s}
\end{equation}
Similar computations can be done for the form factors of other local operators. See section 6.2 of \cite{He:2024nwd}. Such behavior is the leading behavior at asymptotically large energies. In the current GTB implementation, this behavior is implemented by requiring that, above a certain given energy $\sqrt{s_0}\sim 2\text{ GeV}$, the form factor is bounded by
\begin{subequations}\label{chiIL}
\begin{eqnarray}
      |F_0^0(s>s_0)| &\le& \chi^0_0 {m_\pi^4 \over s} \ln \left( {s \over m_\pi^2} \right) \\
  |F_{1}^1(s>s_0)| &\le& \chi_1^1 {16 \pi \alpha_s f_\pi^2 \over s}\\
    |F_2^0(s>s_0)| &\le& \chi_2^0 { 48 \pi \alpha_sf_\pi^2 \over s}
\end{eqnarray}
\end{subequations}
where a numerical factor $\chi^I_\ell$ of $O(1)$ was introduced. Such factors are a simplified way to account for subleading corrections present at $\sqrt{s} \gtrsim 2\,\GeV$.  
As studied in \cite{He:2025gws}, the value of such multipliers affects the low energy dynamics quantitatively.\footnote{Keeping $\chi^I_\ell$ of order $O(1)$ does not change qualitative aspects such as the existence of resonances.}

Finally, at very low energies, the pion amplitude is well approximated by the tree-level expression:
\begin{equation}\label{treeA}
    A_{\text{tree}}(s,t,u) = {4 \over \pi}{ s-m_\pi^2 \over 32 \pi f_\pi^2}
\end{equation}
leading to only three non-vanishing partial waves:
\begin{equation}
\begin{array}{l}
    f^0_{0,\text{tree}}(s)= {2 \over \pi}{2s - m_\pi^2 \over 32 \pi f_\pi^2},\;\; f^1_{1,\text{tree}}(s)= {2 \over \pi}{s - 4 m_\pi^2 \over 96 \pi f_\pi^2},\;\;
    f^2_{0,\text{tree}}(s)= {2 \over \pi}{2 m_\pi^2-s  \over 32 \pi f_\pi^2}
    \end{array}
\end{equation}
To constrain the partial waves in the low energy unphysical region $0\le s\le 4$, we follow the procedure proposed in \cite{PhysRevLett.133.191601,PhysRevD.110.096001} and impose the following constraints: 
\begin{equation}\label{chimatch}
\begin{aligned}
    &||[f^0_0(s_j)-R^\text{tree}_{01}(s_j)f^1_1(s_j),f^2_0(s_j)-R^\text{tree}_{21}(s_j)f^1_1(s_j)]||\le \epsilon^\chi\\
    &R^\text{tree}_{01}=\frac{f^0_{0,\text{tree}}}{f^1_{1,\text{tree}}},\;R^\text{tree}_{21}=\frac{f^2_{0,\text{tree}}}{f^1_{1,\text{tree}}},\;s_j=\{1/2,1,3/2,2\}
\end{aligned}
\end{equation}
to match partial waves with the tree-level ones up to a small tolerance $\epsilon^{\chi}$.

\bigskip

Thus we have all the generic and dynamical constraints that the bootstrap variables
\begin{equation}\label{para}
    \{T_0, \sigma_1,\sigma_2,\rho_1, \rho_2,\text{Im} F^I_\ell, \rho_\ell^I\}
\end{equation}
have to satisfy. To find the right solution to this set of constraints, \cite{He:2025gws} introduced a so-called Watsonian iteration procedure. The idea is that two pion states dominate at low energies \footnote{This is the case even beyond the inelastic threshold.} and the positive semidefinite matrix \eqref{Bmt} corresponding to such a physical solution should fully saturate the positivity constraints and thus have two zero modes. The saturation condition has three important consequences: no particle production at this energy, Watson's theorem that relates the phase of the form factor with the phase shift and the saturation of the spectral density by the 2-pion form factor:
\begin{equation}\label{saturation}
1+ih^I_\ell(s)=S^I_\ell(s)=e^{2i\delta_\ell(s)} = \frac{F^I_\ell(s)}{F^{I*}_\ell(s)},\ \  \rho^I_\ell(s)=|\mathcal{F}^I_\ell(s)|^2
\end{equation}
These equations are nonlinear in the bootstrap variables \eqref{para} and thus hard to solve. An efficient way to find solutions saturating the constraints is to start at a given point of the parameter space, and iteratively maximize linear functionals that push the system towards the saturation conditions \eqref{saturation}. Such functionals are respectively given by:
\begin{equation}\label{F1}
    \begin{aligned}
       \mathcal{F}_1=\int ds\sum_{I,\ell}\left\{\frac{|h^I_\ell|}{|F^I_\ell|}\bigg|_{\text{old}}(\text{Re}F^I_{\ell}|_{\text{old}}\text{Re}h^I_{\ell}|_{\text{new}}+\text{Im}F^I_{\ell}|_{\text{old}}\text{Im}h^I_{\ell}|_{\text{new}})-\text{Im}h^I_{\ell}|_{\text{new}}\right\}
    \end{aligned}
\end{equation}
and
\begin{equation}\label{F2}
    \mathcal{F}_2=\int ds\sum_{I,\ell}\left\{2\text{Re}\mathcal{F}^I_{\ell}|_{\text{old}}\text{Re}\mathcal{F}^I_{\ell}|_{\text{new}}+2\text{Im}\mathcal{F}^I_{\ell}|_{\text{old}}\text{Im}\mathcal{F}^I_{\ell}|_{\text{new}}-\rho^I_\ell|_{\text{new}}\right\}
\end{equation}
The subscript $_{\text{old}}$ denotes the previous solution and $_{\text{new}}$ the one being optimized. The main point is that an ``old" solution \emph{ that satisfies} \eqref{saturation} gives, after some simple algebra, and discarding terms that do not depend on the ``new" solution, functionals:
\beqa
 \mathcal{\tilde{F}}_1 &=& \half \int ds\sum_{I,\ell}\ |S_\ell^I(s)_{\text{new}} |^2 - \left|S_\ell^I(s)_{\text{new}}- S_\ell^I(s)_{\text{old}}\right|^2 \label{FT1}\\
 \mathcal{\tilde{F}}_2 &=&       \int ds\sum_{I,\ell}\ |\cF_\ell^I(s)_{\text{new}} |^2 -\rho_\ell^I(s)_{\text{new}} - \left|\cF_\ell^I(s)_{\text{new}}- \cF_\ell^I(s)_{\text{old}}\right|^2 \label{FT2}
\eeqa
% \yifei{i am not sure how you wrote this, it is a bit confusing: for the second line, $|\mathcal{F}|^2$? for the first line, it looks like pure unitarity saturation and there is no form factor?} \mk{There is no form factor (new) in the original functional}\yifei{but old form factor has to show up otherwise it is not watson} \mk{These formulas are the reason we use these functionals, if you do the calculation I say before \eqref{FT1} you get what\eqref{FT1}, \eqref{FT2}, or not? It proves the saturated solutions are fixed points of iteration as said below} 
Furthermore, the maximum of $ \mathcal{\tilde{F}}_{1,2}$ with respect to the new solution is given by the new solution being equal to the old one. Therefore, solutions that satisfy \eqref{saturation} are fixed points of the iteration procedure. Thus, iterating the maximization procedure until convergence gives the solution we are interested in. See section 2.2 of \cite{He:2025gws} for more details.

\section{Varying $N_c$}

The above procedure was previously carried out for $N_c=3$ and led to a converged result in good agreement with experiment \cite{He:2025gws}. To study gauge theories of different colors, we vary $N_c$ while keeping the t'Hooft coupling fixed \cite{tHooft:1973alw}:
\begin{equation}\label{thooft}
    g_{\text{YM}}^2N_c\sim \alpha_s N_c\;\;\text{fixed}
\end{equation}
which should result in a well-defined $1/N_c$ expansion of the results. Since $N_c$ is a gauge theory parameter, its value appears directly in the UV information as can be seen in eqs. \eqref{PIS}-\eqref{PIV}. They show that the sum rules are proportional to $N_c$. In practice, we use the higher results in Appendix \ref{higherloops}, but, in the 't Hooft limit they are also proportional to $N_c$. For the spin-2 channel, instead of the energy momentum tensor whose sum rules grow as $N_c^2$, we take its quark part:
\begin{equation}\label{jq}
j_T=\bar{\psi}_A\slashed{\Delta}D_{\Delta}\psi_A
\end{equation}
The main motivation for this choice is that its 2-point function grows as $N_c$ as seen in eq. \eqref{2pt} below. Thus, the sum rules in all channels behave in the same way as we increase $N_c$. From the physical point of view, as $N_c$ increases, the $f_2$ meson in this channel should be better described as a $q\bar{q}$ state created by the quark part of $T_{\mu\nu}$. One caveat of using the operator \eqref{jq} is that it mixes with the gluon part of the energy momentum tensor and therefore it is scale ($\mu$) and scheme dependent. The operator defined at a scale $\mu$, is related to the operator at a scale $\mu_0$ in such a way that the form factor at zero momentum transfer satisfies \cite{Belitsky:2005qn,Hatta:2018sqd,Stegeman:2025tdl}:
\beq
F_q(t=0, \mu) = \frac{N_f}{4C_F+N_f} +  \left[F_q(t=0,\mu_0)- \frac{N_f}{4C_F+N_f}\right]\, \left(\frac{\alpha_s(\mu)}{\alpha_s(\mu_0)}\right)^{\frac{8C_F+2N_f}{11N_c-2N_f}} \label{RGF}
\eeq
where $C_F=\frac{C_2}{N_c}$, $C_2=\frac{N_c^2-1}{2}$. At very high energy $\mu\rightarrow\infty$, $\alpha_s\rightarrow 0$ and
\beq
F_q(t=0, \mu\rightarrow\infty) = \frac{N_f}{4C_F+N_f} \label{Fq0}
\eeq
However, in our computation, we take the operator at a scale $\mu=\sqrt{s_0}\sim 2\text{GeV}$. The situation is similar to the asymptotic form factors where the theoretical calculation is valid at very large energies. In this paper we use, as an initial approximation, the asymptotic normalization \eqref{Fq0} corrected  by a factor that accounts for differences that are known to exist from lattice\footnote{Lattice calculations \cite{Hackett:2023nkr} for $N_c=3$, give the $u+d$ quark contribution to be $F_q(0)\simeq 0.46$ close but not equal to the asymptotic value \eqref{Fq0} $F_q(0) \simeq 0.27$. Also, for large $N_c$ \eqref{Fq0} behaves as $N_c^{-1}$ and the main contribution would come from the second term in \eqref{RGF} which should have a finite value. However we only go up to $N_c=12$ } results \cite{Hackett:2023nkr}. Notice also that for very large $N_c$, \eqref{Fq0} behaves as $N_c^{-1}$ and the main contribution to $F_q(0)$ would come from the second term in \eqref{RGF} which should be taken into account if one wants to extend the range of $N_c$ further. 

Finally, the two point functions of the operators \eqref{jq} are also known \cite{narison_2004} and can be found in Appendix \ref{higherloops}. 

Another place $N_c$ appears is in the asymptotic form factor which, for the vector and tensor current take the form \cite{He:2024nwd}:\footnote{For the scalar form factor whose asymptotic is not known, we take $F^0_0(t)\overset{t\to \infty}{\simeq}\frac{m_\pi^4}{t}\ln\left(\frac{t}{m_\pi^2}\right)$ considered in \cite{He:2024nwd} based on dimensional analysis.}
\begin{subequations}
\begin{eqnarray}
    F_1^1(t)&\overset{t\to\infty}{\simeq}&-\frac{36\pi\alpha_sf_\pi^2}{t}\frac{C_2(N_c)}{N_c^2}\\
    F_0^2(t)&\overset{t\to\infty}{\simeq}&-\frac{84i\pi\alpha_sf_\pi^2}{t}\frac{C_2(N_c)}{N_c^2}
    \end{eqnarray}
\end{subequations}
Note that the explicit $N_c$ dependent factor $\frac{C_2(N_c)}{N_c^2}$ approaches one as $N_c$ gets larger. The asymptotic form factor also has an implicit dependence on $N_c$ since, at leading order 
\begin{equation}
    f_\pi\sim\sqrt{N_c}
\end{equation}
as follows from its definition through the axial current matrix element.
Since we keep the t'Hooft coupling \eqref{thooft} fixed, $\alpha_sf_\pi^2\sim O(1)$ and the asymptotic form factors stay approximately constant as $N_c$ increases.

Finally, as described before, for controlling the very low energy region, we introduced a tolerance $\epsilon^{\chi}$ \eqref{chimatch} that constrains the allowed deviation from the two derivative action (Weinberg model). Since derivatives do not affect the $N_c$ counting, higher derivative corrections will be of the same order as two derivatives, namely $\epsilon^\chi\sim 1/N_c$. We choose:
\begin{equation}\label{echiNc}
    2\times 10^{-3}\frac{3}{N_c} \le \epsilon^{\chi} \le 4\times 10^{-3}\frac{3}{N_c}
\end{equation}
In other words, as $N_c$ increases, pion dynamics is given by the tree-level action, but the tree level action contains arbitrary number of derivatives. The derivative corrections increase with energy and become of order one at the mass of the first resonance which therefore is approximately independent of $N_c$.  

Overall, it is straight-forward to setup the problem for various values of $N_c$ by simply changing the numerical values of the inputs. Below, we present results for $3\le N_c\le12$ since beyond that value we do not get a good convergence for the solutions anymore, probably because the resonances become too narrow. Notice that the resonances become narrow since the low energy coupling decreases as $N_c$ increases. 

\subsection{Discussion of errors}\label{err}

The procedure has two main sources of errors. One is due to the high energy inputs, in particular the asymptotic form factors in \eqref{chiIL}. To estimate this error we explore a range of values for the coefficients $\chi$ in \eqref{chiIL}. Choosing several values for $\chi$ in the range:
\beq\label{chirange}
2\le\chi_{S0}\le 10, \ \ \ 1.5 \le \chi_{P1} \le 3, \ \ \ \ 2.5 \le \chi_{D0} \le 5,
\eeq
we obtain a set of results. The other kind of error is systematic and depends on the details of the implementation of the procedure. For example, in one case we used an initial functional that, for each $N_c$ maximizes $\lambda$, the pion coupling defined in \eqref{def_lambda} below. In another case we used as a starting point an $N_c=3$ solution obtained using no functional that gave a lower value of $\lambda$. The first procedure systematically overestimated $\lambda$ for every $N_c$ (\eg\ $\lambda\simeq 0.027$ for $N_c=3$) whereas the other underestimated it ($\lambda\simeq 0.017$ for $N_c=3$) compared with the Weinberg model $\lambda \simeq 0.023$.  Using different variations including changing the asymptotic form factors in the range \eqref{chirange} we obtained a set of 35 solutions. For each quantity we plot the average over those solutions and include $2\sigma$ error bars that should contain most of the points. Thus, the meaning of the error is not statistical but a region which contains most of our results.

\section{Low energy $\chi$PT region}\label{LEC}

The results in the low energy region are put, as is standard, in the framework of chiral perturbation theory ($\chi$PT) \cite{Gasser:1984gg,GASSER198765}. The low energy region, where derivative corrections are small, includes the Mandelstam triangle in the unphysical region and the near threshold region below the first resonances in the physical region. The size of this region is basically independent of $N_c$.  
For comparison with experiment, the main parameters are the scattering lengths, effective range and charge radius of the pion in the various channels. From a bootstrap perspective, the main quantity is the scattering amplitude that can be characterized by the expansion coefficients around the symmetric point $s=t=u=\frac{4}{3}\mpi^2$. Finally we derive the renormalized $\chi$PT low energy constants (LEC) $\bar{\ell}_j$ useful to compare with the phenomenological literature.  In the rest of this section we use units such that $\mpi=1$. 

\subsection{Scattering lengths, effective range parameters and charge radii}

 Defining 
\beq
S^I_\ell(s) = 1 + i h^I_\ell(s) 
\eeq
below the four pion threshold, where there is no particle production, we have
\begin{equation}
    S^I_\ell = e^{2i\delta^I_\ell(s)},\;h^I_\ell=2\sin\delta^I_\ell\, e^{i\delta^I_\ell}
\end{equation}
Near threshold we can expand the partial waves as:
\beq
\Re\, h^I_\ell\ \simeq_{k\rightarrow 0} \frac{4}{\sqrt{s}}\, k^{2\ell+1} \left(a^I_\ell+b^I_\ell k^2 + \ldots\right), \ \ \ k=\sqrt{\frac{s}{4}-1}
\eeq
that defines the scattering lengths $a^I_\ell$ and effective range $b^I_\ell$ in the standard manner. 

The pion charge radii are defined through the low energy expansion of the form factor as:
\begin{equation}
F^I_\ell(s)=F^I_\ell(0)\left(1+\frac{1}{6}s\langle r^2\rangle^I_\ell+\ldots\right)
\end{equation}
We can then extract the scalar, vector and tensor radius\footnote{In this paper we consider the tensor radius as the one measured by the quark part of the energy momentum tensor \eqref{jq}. Note that unlike the normalizations of the scalar and vector form factors \eqref{f0}, the tensor form factor normalization changes with $N_c$.} from the corresponding form factors. 

\begin{figure}[h]
	\centering
	\includegraphics[width=1.\textwidth]{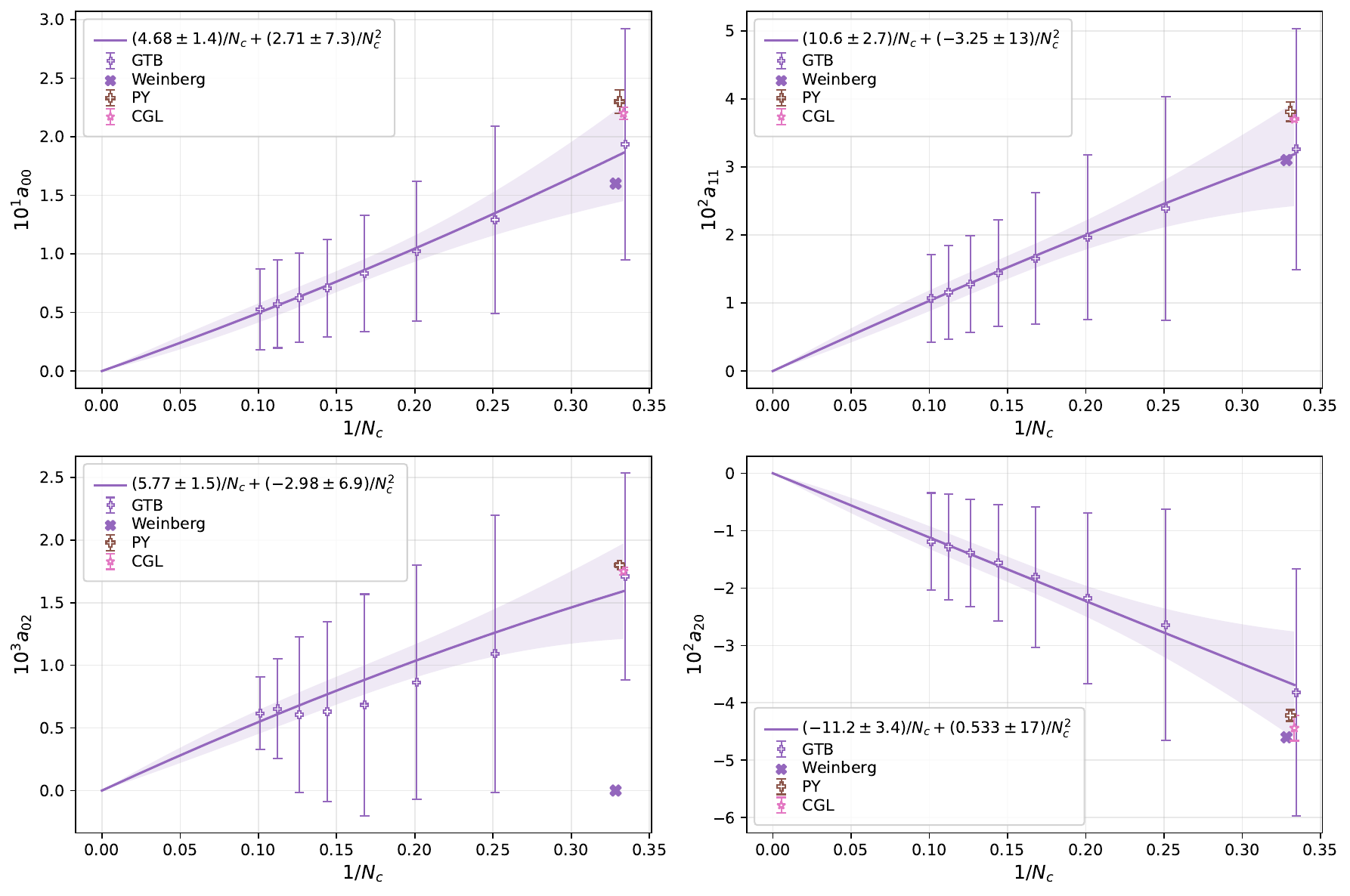}
	\caption{Scattering lengths $a^I_\ell$ as functions of $N_c^{-1}$.  We present the averaged results  with $2\sigma$ error bars and the fitting functions with a $1\sigma$ band. The $N_c=3$ values are compared with Weinberg model \cite{PhysRevLett.17.616} and phenomenological determinations {\bf PY} \cite{Pelaez:2004vs} and {\bf CGL}\cite{Colangelo:2001df}. }
	\label{scatlen}
\end{figure}

\begin{figure}[h]
	\centering
	\includegraphics[width=1.\textwidth]{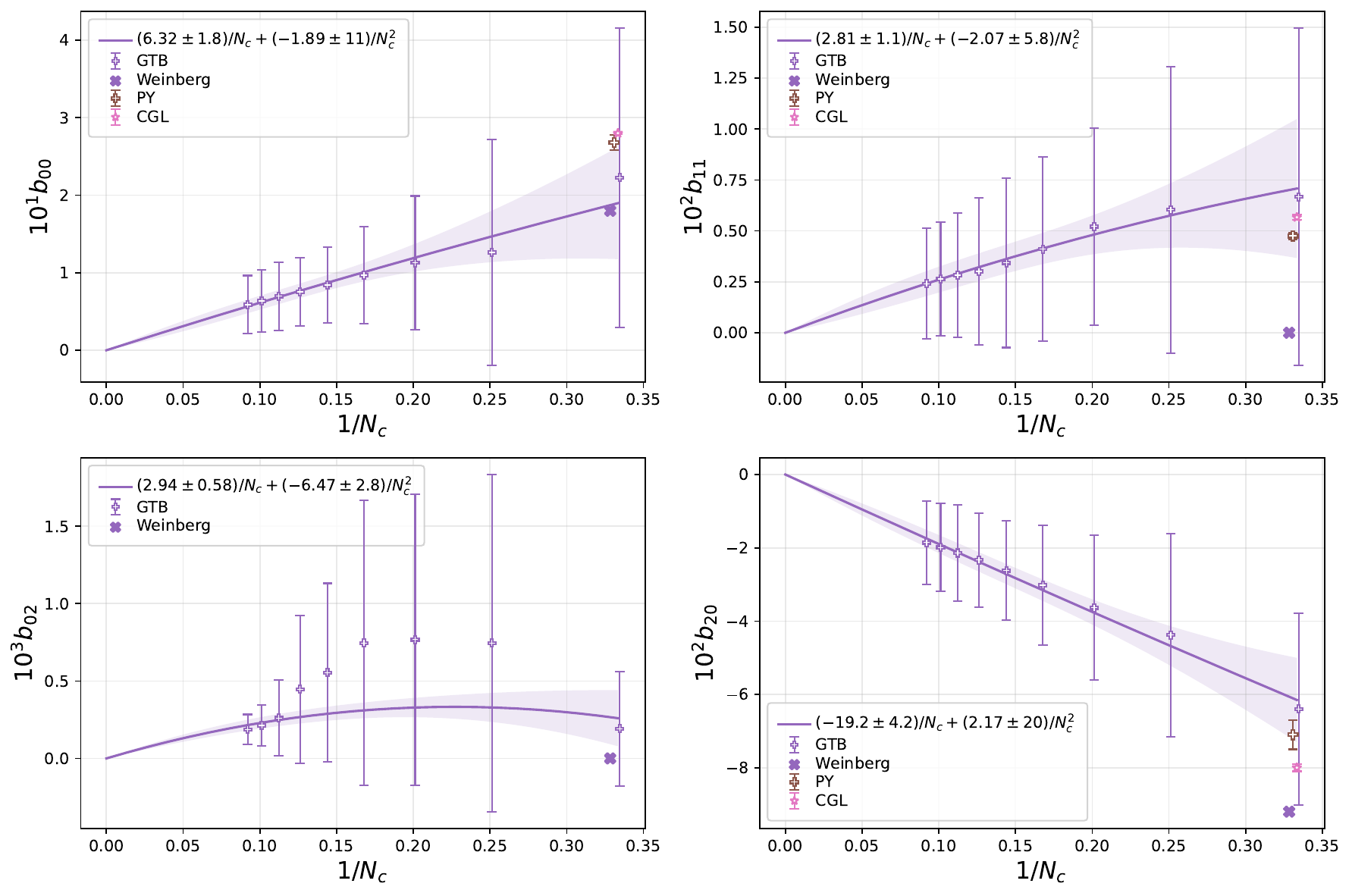}
	\caption{Effective range $b^I_\ell$ with varying $N_c$.  We present the averaged results  with $2\sigma$ error bars and the fitting functions with a $1\sigma$ band. The $N_c=3$ values can be compared with Weinberg model \cite{PhysRevLett.17.616} and phenomenological determinations {\bf PY} \cite{Pelaez:2004vs} and {\bf CGL}\cite{Colangelo:2001df}.}
	\label{effrange}
\end{figure}

\begin{figure}[h]
	\centering
	\includegraphics[width=0.8\textwidth]{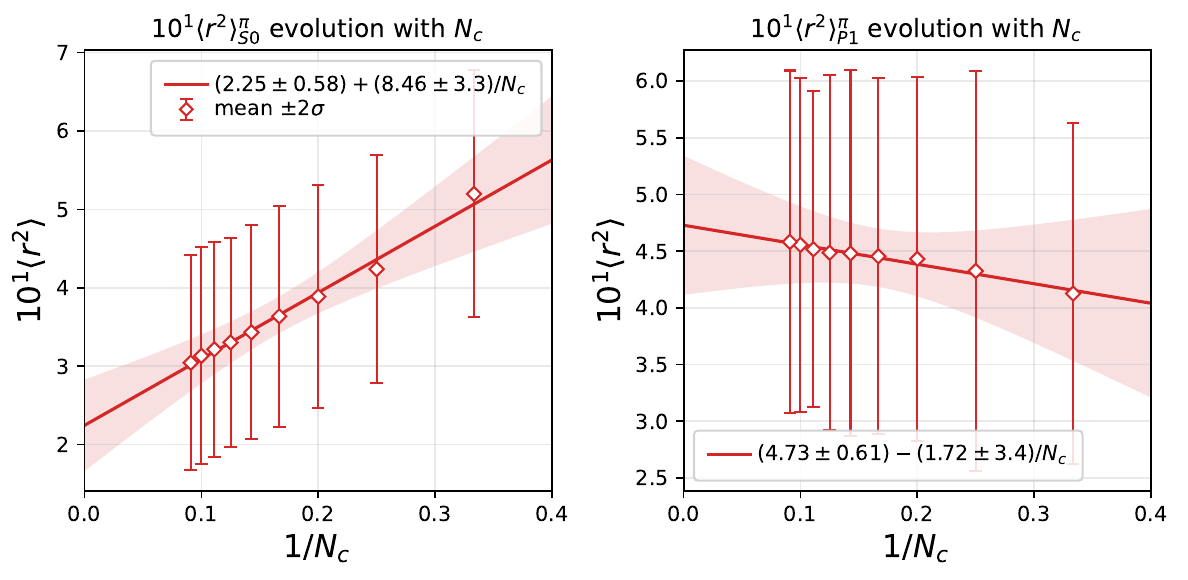}
	\caption{Pion radius $\langle r\rangle^I_\ell$ with varying $N_c$.  We present the averaged results  with $2\sigma$ error bars and the fitting functions with a $1\sigma$ band. }
	\label{rpi}
\end{figure}

In Figs. \ref{scatlen}, \ref{effrange} and \ref{rpi}, we show the results for the scattering lengths $a^I_\ell$, effective range $b^I_\ell$ and charge radii $\langle r\rangle^I_\ell$ with varying $N_c$ using $\mpi=1$ units.

%\begin{eqnarray}
%         a^{(0)}_0&=&7 \lambda +\frac{\lambda ^2}{9 \pi }  \left[64 \bar{\ell}_1+128 \bar{\ell}_2+144 \bar{\ell}_3+84+231 \sqrt{2} \tan ^{-1}(2
%   \sqrt{2})\right]+\mathcal{O}(\lambda^3)\nonumber\\
%   a^{(2)}_0&=&-2 \lambda +\frac{2\lambda ^2}{9 \pi } \left[32 \bar{\ell}_1+64 \bar{\ell}_2-36 \bar{\ell}_3+42-33 \sqrt{2} \tan ^{-1}(2 \sqrt{2})\right]+\mathcal{O}(\lambda^3)\nonumber\\
%   a^{(1)}_1&=&\frac{4 \lambda }{3}+\frac{2 \lambda ^2}{27 \pi }\left[-64 \bar{\ell}_1+16 \bar{\ell}_2+54 \bar{\ell}_3-131+66 \sqrt{2} \tan ^{-1}(2
%   \sqrt{2})\right]+\mathcal{O}(\lambda^3)\nonumber\\
%   a_2^{(0)} &=& \frac{32\lambda^2}{45\pi}\left[\bar{\ell}_1+4\bar{\ell}_2-\frac{53}{8}\right]+\mathcal{O}(\lambda^3) \nonumber\\
%a_2^{(2)} &=& \frac{32\lambda^2}{45\pi}\left[\bar{\ell}_1+\bar{\ell}_2-\frac{103}{40}\right]+\mathcal{O}(\lambda^3)\label{larelation}
%\end{eqnarray}

\subsection{Expansion coefficients of the amplitude}

\begin{figure}[h!]
	\centering
	\includegraphics[width=\textwidth]{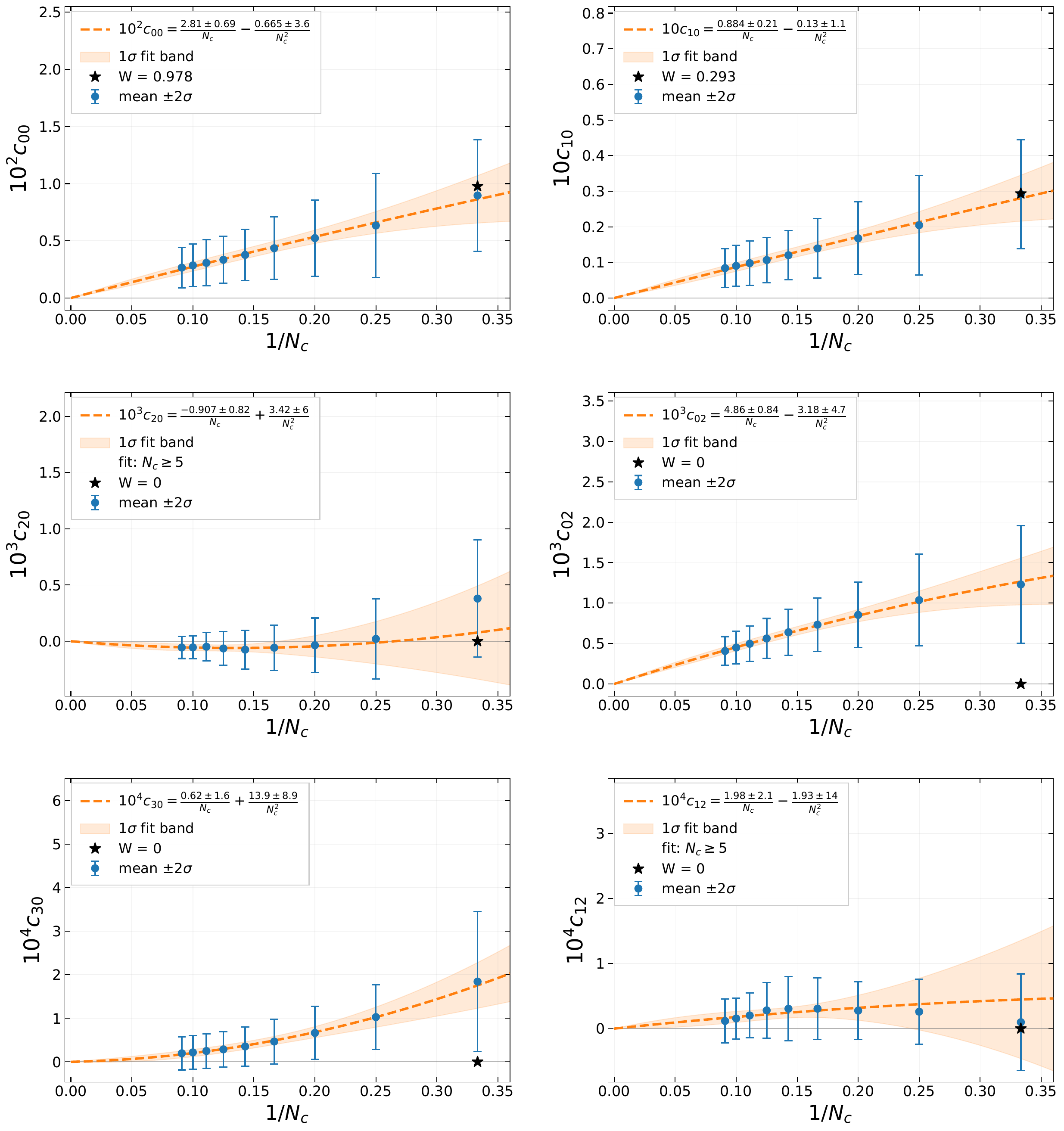}
	\caption{Expansion coefficients $c_{ij}$ with varying $N_c$. We present the averaged results  with $2\sigma$ error bars and the fitting functions with a $1\sigma$ band. The $N_c=3$ values are compared with Weinberg model {\bf W} \cite{PhysRevLett.17.616}.}
	\label{cmn}
\end{figure}

The pion scattering amplitude plays a central role in the bootstrap and is completely determined by the function $A(s,t,u)=A(s,u,t)$ defined in \eqref{amp}. It can be expanded around the symmetric point $s=t=u=\frac{4}{3}$ (using units where $\mpi=1$) in terms of the variables:
\beq\bar{s} = s-\frac{4}{3}, \ \ \ \ \nu=\frac{t-u}{2}
\eeq
as:
\beq\label{Asymexp}
A(s,t,u) = c_{00} + c_{10}\, \bar{s} + c_{20}\, \bar{s}^2 + c_{02}\, \nu^2 + c_{30}\, \bar{s}^3 + c_{12}\, \bar{s}\,\nu^2 + \ldots 
\eeq
where only even powers of $\nu$ appear on account of the symmetry $A(s,t,u)=A(s,u,t)$. In fig. \ref{cmn}, we show the results of the expansion coefficients for varying $N_c$. 
Of particular importance is the pion coupling $\lambda$ defined as:
\begin{equation} \label{def_lambda}
    \lambda=\frac{1}{32\pi}\mathcal{M}(\pi^0\pi^0\to\pi^0\pi^0)|_{s=t=u=\frac{4}{3}m_\pi^2}=\frac{3\pi}{4}A\bigg(\frac{4}{3}m_\pi^2,\frac{4}{3}m_\pi^2,\frac{4}{3}m_\pi^2\bigg)=\frac{3\pi}{4}\, c_{00}
\end{equation}
Also, from the expansion \eqref{Asymexp} we can compute the partial waves $f^I_\ell(s)$ and then the scattering length and effective range obtaining the expressions in appendix \ref{LEformula}.  
 The results are consistent with the direct evaluations shown in figs. \ref{scatlen} and \ref{effrange}.

\subsection{Low energy constants $\bar{\ell}_j$ (Wilson coefficients)}

\begin{figure}[h!]
	\centering
	\includegraphics[width=0.94\textwidth]{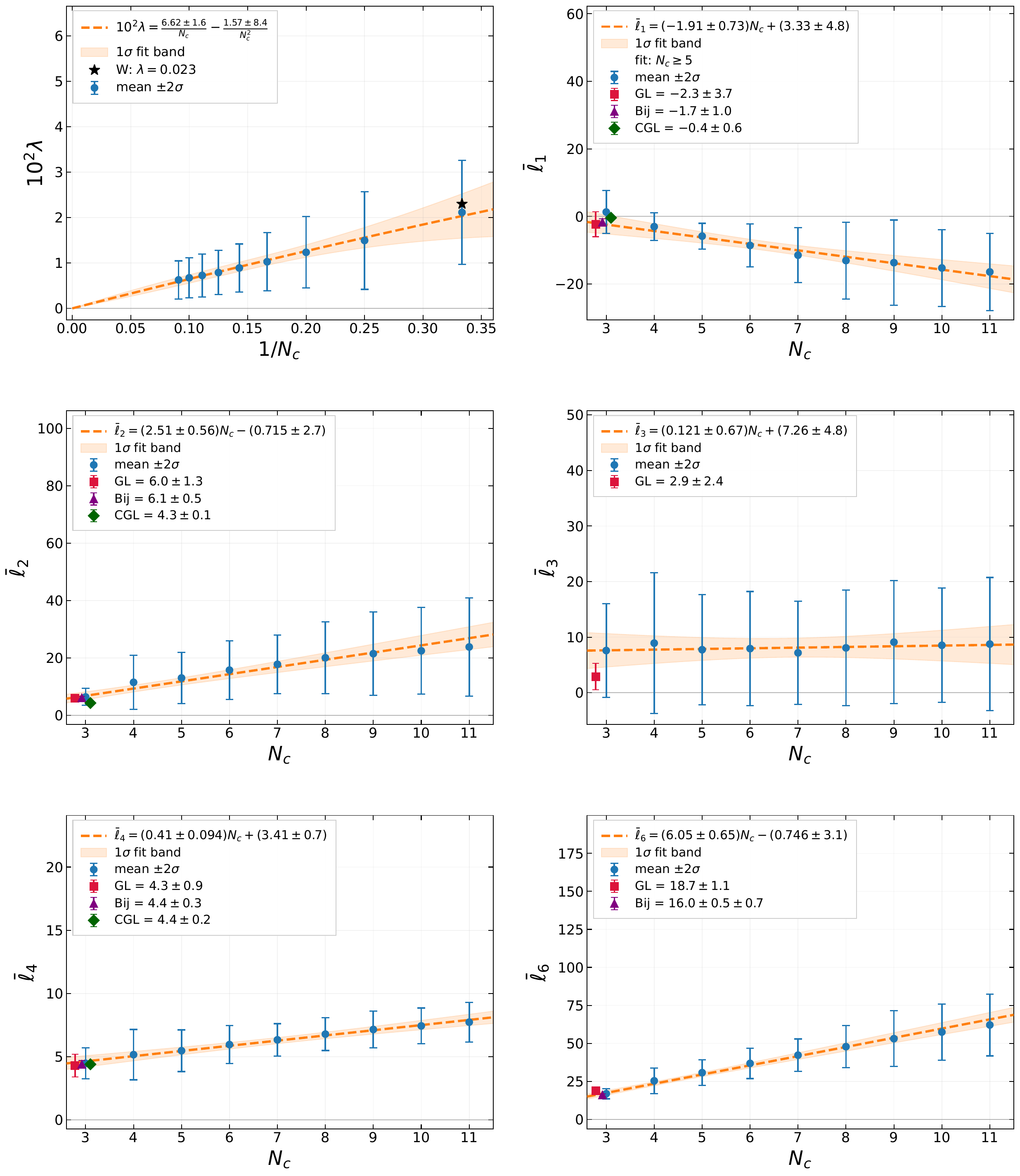}
	\caption{Low energy constants with varying $N_c$. We present the averaged results  with $2\sigma$ error bars and the fitting functions with a $1\sigma$ band. The $N_c=3$ values are compared with phenomenological determinations {\bf GL} \cite{GASSER1984142}, {\bf Bij} \cite{Bijnens:1994ie,Bijnens:1997vq,Bijnens:1998fm} and {\bf CGL}\cite{Colangelo:2001df}.}
	\label{ells}
\end{figure}

The low energy constants are the couplings of the chiral effective field theory. In \cite{GASSER1984142}, these are defined by choosing the renormalization point to be the pion mass. They can be related to the amplitude expansion and pion radii as:
\beqa
\bar{\ell}_1 &=&  \frac{3\pi^2}{32\lambda^2} (- c_{02}  + 4c_{20}) + \overline{\overline{\ell}}_1 \\
\bar{\ell}_2 &=& \frac{3\pi^2}{16 \lambda^2} c_{02} + \overline{\overline{\ell}}_2\\
\bar{\ell}_3 &=& \frac{\pi^2}{36 \lambda^2} (3 c_{10} + 8 c_{20})  - \frac{\pi}{3\lambda} \overline{\overline{\ell}}_3\\
\bar{\ell}_4 &=& \frac{13}{12} + \frac{\pi}{12\lambda} \langle r^2\rangle^0_0  \\
\bar{\ell}_6 &=& 1 + \frac{\pi}{2\lambda} \langle r^2\rangle^1_1 
\eeqa
where
\beqa
\overline{\overline{\ell}}_1 &=&  -\frac{1733}{1024} \sqrt{2}\arctan(2\sqrt{2})) + \frac{2119}{768} \simeq -0.187 \\
\overline{\overline{\ell}}_2 &=&  \frac{1037}{512} \sqrt{2}\arctan(2\sqrt{2}))  - \frac{991}{384}  \simeq 0.945\\
\overline{\overline{\ell}}_3 &=& -\frac{17}{16}\sqrt{2}\arctan(2\sqrt{2})) + \frac{25}{12}  \simeq 0.234
\eeqa
These constants provide an equivalent descriptions of the low energy physics and can be compared directly with known phenomenological results. We show in fig. \ref{ells} the results for these.

\subsection{Summary of low energy results (in $\mpi$ units)}\label{summary_low_energy}

 To summarize, we obtained, as function of the number of colors $N_c$ the scattering lengths and effective range:
 \begin{equation}
\begin{aligned}
a^0_0 &=\ \ (4.68\pm 1.4)  \times 10^{-1}\, \frac{1}{N_c}, &b^0_0 &=\ \ (6.32\pm 1.8)  \times 10^{-1}\, \frac{1}{N_c}, \\
a^1_1 &=\ \ (10.6\pm 2.7)  \times 10^{-2}\, \frac{1}{N_c}, &b^1_1 &=\ \ (2.81\pm 1.1)   \times 10^{-2}\, \frac{1}{N_c}, \\
a^0_2 &=\ \ (5.77\pm 1.5)  \times 10^{-3}\, \frac{1}{N_c}, &b^0_2 &=   (2.94\pm 0.58) \times 10^{-3}\, \frac{1}{N_c}, \\
a^2_0 &=   -(11.2\pm 3.4)  \times 10^{-2}\, \frac{1}{N_c}, &b^2_0 &=   -(19.2\pm 4.2)    \times 10^{-2}\, \frac{1}{N_c},
\end{aligned}
\end{equation}
% \beqa
% c_{00} &=& \left[(2.8\pm 0.7)\frac{1}{N_c} - (0.7\pm 3.6)\frac{1}{N_c^2}\right]\times 10^{-2} \\
% c_{10} &=& \left[(0.9\pm 0.2)\frac{1}{N_c} - (0.1\pm 1)\frac{1}{N_c^2}\right]  \times 10^{-1} \\
% c_{20} &=& -\left[(0.9\pm 0.8)\frac{1}{N_c} - (3\pm 6)\frac{1}{N_c^2}\right]   \times 10^{-3} \\
% c_{02} &=& \left[(4.9\pm 0.8)\frac{1}{N_c} - (3.2\pm 5)\frac{1}{N_c^2}\right]  \times 10^{-3} \\ 
% c_{30} &=& \left[(0.62\pm 1.6)\frac{1}{N_c} - (14\pm 9)\frac{1}{N_c^2}\right]  \times 10^{-4} \\
% c_{12} &=& \left[(1.98\pm 2.1)\frac{1}{N_c} - (2\pm 14)\frac{1}{N_c^2}\right]  \times 10^{-4} 
% \eeqa
the coefficients of the expansion of the amplitude $A(s,t,u)$ around the symmetric points \eqref{Asymexp}: 
\begin{align}
 c_{00} &=\ \ (2.8\pm 0.7)  \times 10^{-2}\, \frac{1}{N_c},\;\;c_{10} =\ \ (0.9\pm 0.2)  \times 10^{-1}\, \frac{1}{N_c}\\
 c_{20} &= -(0.9\pm 0.8)    \times 10^{-3}\, \frac{1}{N_c},\;\;c_{02} =\ \ (4.9\pm 0.8)  \times 10^{-3}\, \frac{1}{N_c}\\
 c_{30} &=\ \ (0.62\pm 1.6) \times 10^{-4}\, \frac{1}{N_c},\;\;c_{12} =\ \ (1.98\pm 2.1) \times 10^{-4}\, \frac{1}{N_c}
\end{align}
The errors of the subleading terms $\frac{1}{N_c^2}$ are larger so we only present them in the plots in fig. \ref{cmn}. Finally, it is also interesting to compute the Wilson coefficients, namely the couplings in the effective action, in the limit of large $N_c$. If we define 
\beq
\bar{\ell}_{j,\infty}=\lim_{N_c\rightarrow\infty} \frac{\bar{\ell}_j}{N_c}
\eeq
we find (in $\mpi=1$ units):
\begin{subequations}
\beqa
\bar{\ell}_{1,\infty} & = & -1.91\pm 0.73 \\
\bar{\ell}_{2,\infty} & = &\ \ 2.51\pm 0.56 \\
\bar{\ell}_{3,\infty} & = &\ \ 0.12\pm 0.67 \\
\bar{\ell}_{4,\infty} & = &\ \ 0.41\pm 0.1 \\
\bar{\ell}_{6,\infty} & = &\ \ 6.05\pm 0.65
\eeqa
\end{subequations}
whereas $(\lambda\, N_c) \rightarrow (6.62\pm 1.6) \times 10^{-2}$ as $N_c\rightarrow\infty$. The sub-leading $N_c^{-1}$ terms can be found in fig.\ref{ells}. For example we find
\beq
\frac{\bar{\ell}_{4}}{N_c} = (0.41\pm 0.1 ) + \frac{3.4 \pm 0.7}{N_c} \ \ \ \ \ \mathrm{[GTB]}
\eeq
that can be compared with the ($N_f=2$) lattice result \cite{bonanno2025nonperturbativedeterminationmesonmasses, DeGrand:2023hzz}: 
\beq
\frac{\bar{\ell}_{4}}{N_c} = 0.448(54) + \frac{2.98(38)}{N_c} \ \ \ \ \ \mathrm{[lattice]}
\eeq
Finally, using for example eq.(A.13) in \cite{He:2025gws} one can compute $\fpi$ from $\lambda$ and $\bar{\ell}_j$ and fit the results with:
\beq
\frac{1}{\sqrt{N_c}} \frac{\fpi}{\mpi} = (0.42 \pm 0.04)\, + \frac{0.03\pm 0.2}{N_c}   \ \ \ \ \ \mathrm{[GTB]}
\eeq 
whereas \cite{bonanno2025nonperturbativedeterminationmesonmasses} give (after rescaling using $\frac{1}{\sqrt{8t_0}} = \frac{\sqrt{\sigma}}{1.148} = 388 \MeV$ in that paper)
\beq
\frac{1}{\sqrt{N_c}} \frac{\fpi}{\mpi} = (0.41 \pm 0.01)\, - \frac{0.37\pm 0.06}{N_c}   \ \ \ \ \ \mathrm{[lattice]}
\eeq

\section{Strongly coupled region and resonances}\label{res}

The strongly coupled energy region is the one of energies where chiral perturbation theory in terms of pions is no longer valid but below the perturbative QCD region. That region is better characterized in terms of resonances which appear as peaks in the form factors or spectral densities at physical energies/momentum transfers. Moreover, they appear as poles on the second sheet of the analytic partial waves, form factors and current correlators. 

\begin{figure}[h!]
	\centering
	\includegraphics[width=\textwidth]{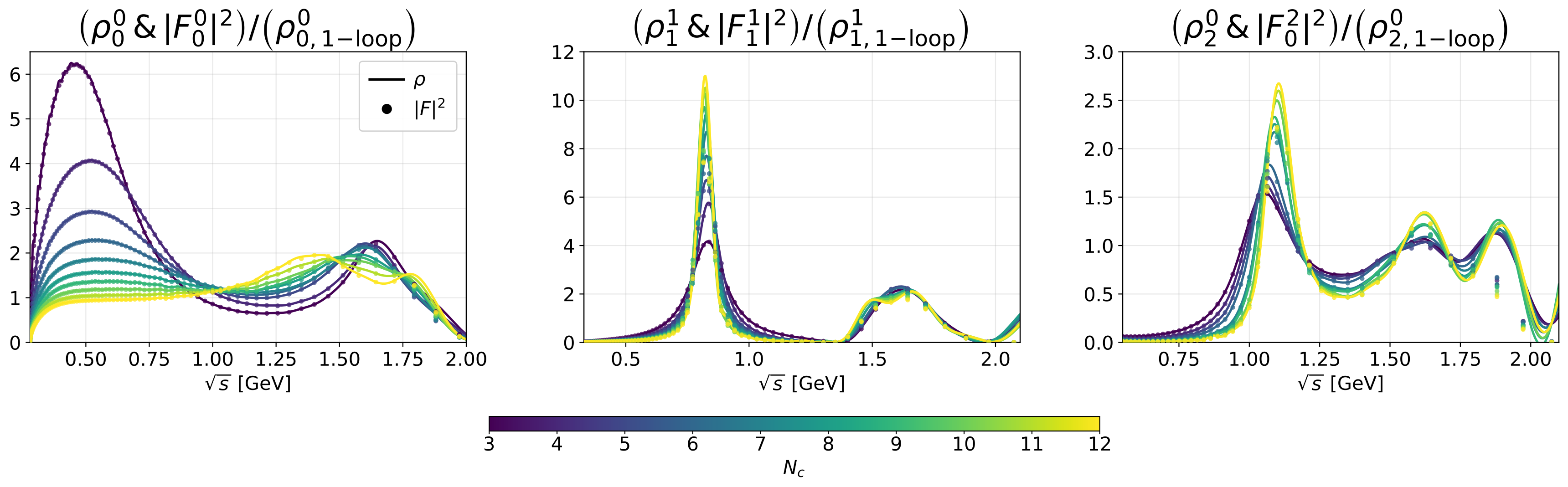}
    \caption{$S0,P1,D0$ spectral densities as a function of energy for various values of $N_c$. We rescale the spectral density with the perturbative QCD 1-loop expression.}
	\label{specS0P1D0}
\end{figure}

In figs. \ref{specS0P1D0} we show an example of the results for the $S0$, $P1$, $D0$ spectral densities, normalized by the one-loop pQCD result and plotted as a function of the energy up to 2 GeV for various values of $N_c$.   
 The $S0$ form factor in fig. \ref{specS0P1D0}, starting at $N_c=3$, has a broad bump around 0.5 GeV which is identified as the $\sigma$ meson or $f_0(500)$ state. Another peak appears at a higher energy and can be identified as the next $f_0$ state. As $N_c$ increases, the $\sigma$ bump gets broader and eventually becomes invisible at higher values of $N_c$.
In the $P1$ spectral density, the first sharp peak at around 0.8 GeV is the signature $\rho(770)$ meson that was a benchmark in the initial GTB implementation at $N_c=3$ \cite{PhysRevLett.133.191601,PhysRevD.110.096001}. With increasing $N_c$, the center of the $\rho(770)$ peak remains mostly unchanged, but the width of the peak gets narrower. This is as expected for the $\rho$ state to be a typical $q\bar{q}$ state. At higher energy, the $P1$ form factor has a broader second peak that was identified as the second state $\rho(1450)$ in the previous $N_c=3$ GTB study. Similar peaks appear in the $D0$ spectral density, where the first peak corresponds to the $f_2$ meson and a second peak also appears at higher energies. The $f_2$ meson, similar to the $\rho$ meson, changes little in energy but gets narrower as $N_c$ increases.

A clearer picture of the resonances in each channel can be seen by extracting the complex poles of $S^I_\ell(s)$ on the second sheet. These poles in the partial waves -- by unitarity saturation in a large range of energies -- become zeros on the first sheet and can be extracted from the analytic amplitudes using the bootstrap parameterization \eqref{amplitude}, \eqref{pw}. The real and imaginary part of the $\sqrt{s_{\text{pole}}}$ are related to the mass $m$ and full-width $\Gamma$ of the resonances as:
\begin{equation}
    \sqrt{s_{\text{poles}}}=m-i\Gamma/2
\end{equation}
Let us now discuss the results for each resonance emphasizing their $N_c$ dependence.

\subsection{$N_c$ dependence of the $\sigma$, $\rho$ and $f_2$ resonances}\label{firstresonance}

\begin{figure}[h!]
	\centering
	\includegraphics[width=\textwidth]{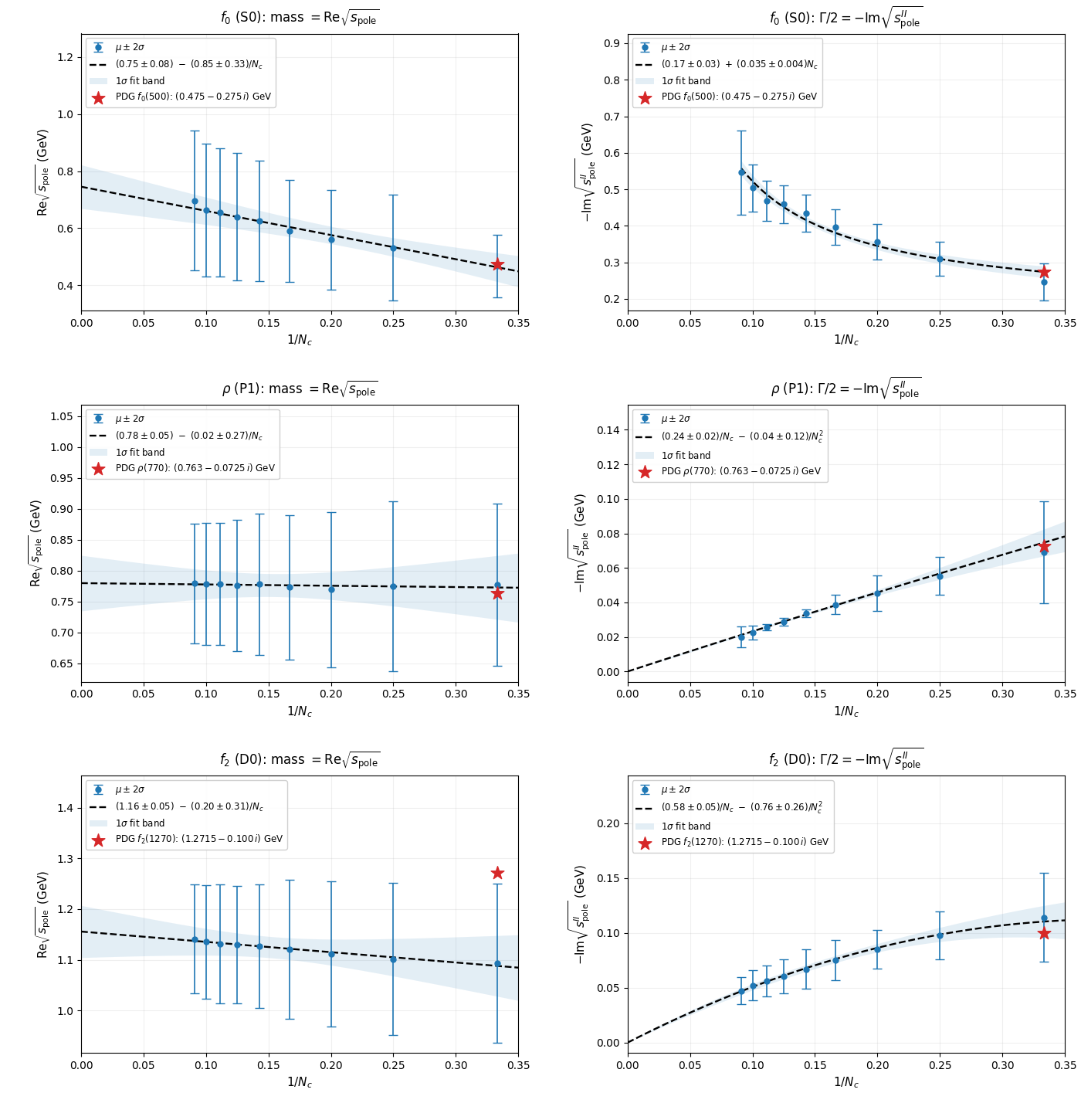}
	\caption{Real and imaginary parts of $\sqrt{s_\text{pole}}$ for the $\sigma$, $\rho$ and $f_2$ meson poles as a function of $N_c^{-1}$. We present the averaged results  with $2\sigma$ error bars and the fitting functions with a $1\sigma$ band. The real part is the mass and the imaginary part is the half-width.}
	\label{poles_Nc}
\end{figure}

The leading resonances in the $S0$, $P1$, and $D0$ channels are respectively the $\sigma$, $\rho$, and $f_2$ mesons. In fig. \ref{poles_Nc}, we plot the average results and the error estimates of the real and imaginary parts of $\sqrt{s_\text{pole}}$ for each of these three resonances. For comparison, we also include the PDG values at $N_c=3$. Overall, we see that for the $\rho$, and $f_2$ resonances, the real parts $\text{Re}\sqrt{s_{\text{pole}}}$ -- the mass -- remains relatively unchanged, as also seen in fig. \ref{specS0P1D0} in peak location. The imaginary parts $\text{Im}\sqrt{s_{\text{pole}}}$ -- half-width -- decrease as $1/N_c$ for increasing $N_c$, namely the poles approach the real axes. This is the expected behavior of $q\bar{q}$ states as corresponds to the $\rho,f_2$ mesons. On the contrary, the $\sigma$ or $f_0$ resonance in the $S0$ channel has an slightly increasing $\text{Re}\sqrt{s_{\text{pole}}}$ (mass) with increasing $N_c$, and, most notably, it gets wider with increasing $N_c$. The increase in the width (see fig. \ref{poles_Nc}) can be fitted with a simple linear function in $N_c$. In previous work \cite{Pelaez:2006nj} using unitarized $\chi$PT it was argued that the sub-leading dependence is $\sqrt{N_c}$ rather than constant, arising naturally from $\sqrt{s_\sigma}\sim f_\pi\sim\sqrt{N_c}$. Our data is not precise enough to determine if such $\sqrt{N_c}$ dependence is present. 

\begin{figure}[h]
	\centering
	\includegraphics[width=\textwidth]{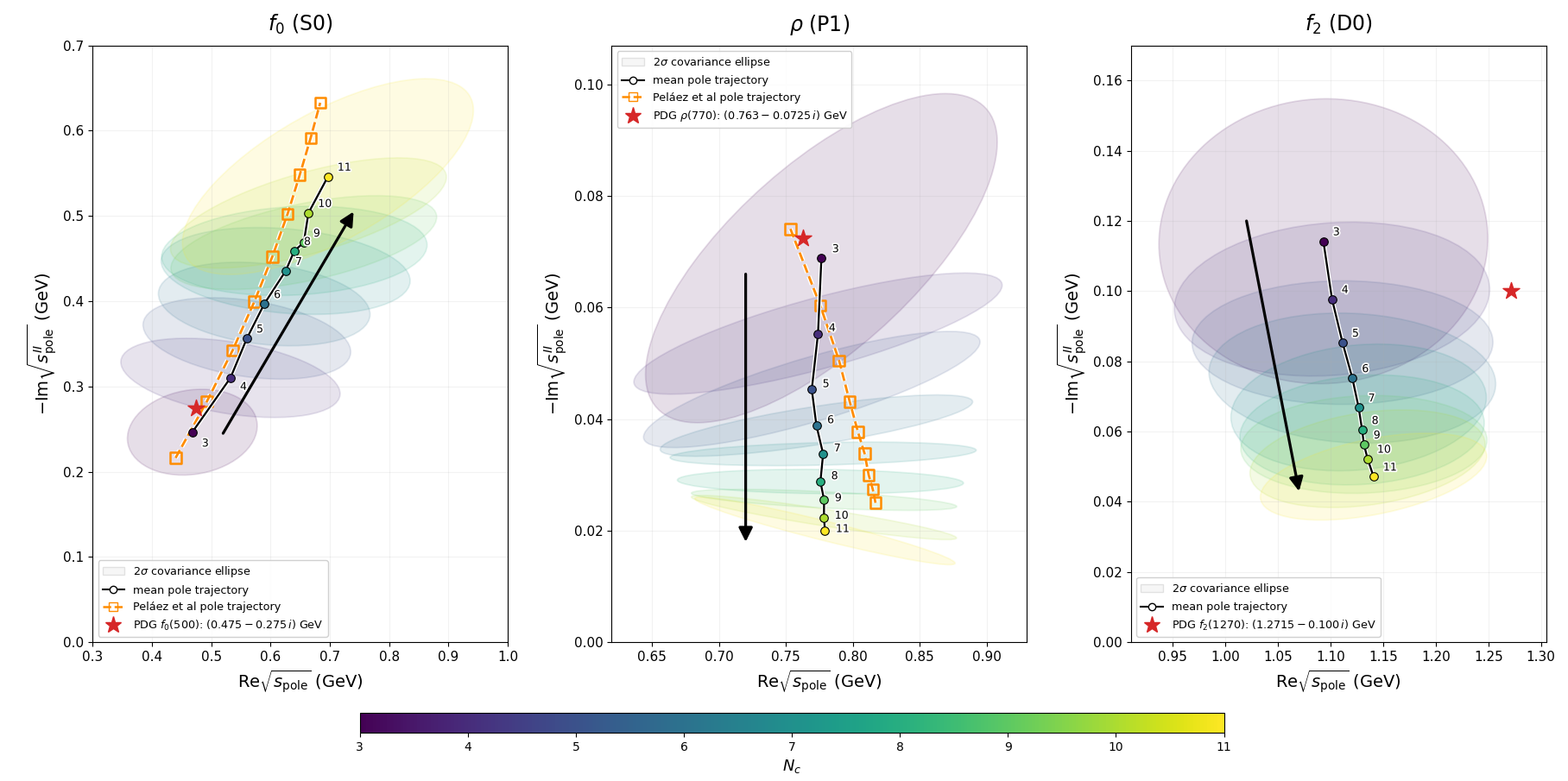}
	\caption{The real and imaginary parts obtained in fig. \ref{poles_Nc} can be used to track the motion of the $\sigma$, $\rho$ and $f_2$ poles in the complex plane. The average is given by the black dots and the ellipse denotes the $2\sigma$ dispersion in the position of the pole. We compare with data from \cite{Pelaez:2006nj} and find agreement within our errors.}
	\label{poles_Nc_s}
\end{figure}

To summarize, we obtain the following $N_c$ dependence of the $\sigma,\rho,f_2$ resonances:
\begin{align}\label{summary_resonances}
        m_\sigma(N_c)&=(0.75\pm 0.08)-(0.85\pm 0.33)\frac{1}{N_c}\;\text{GeV},\\
        \Gamma_\sigma(N_c)&=(0.34\pm 0.06)+(0.07\pm 0.008)N_c\;\text{GeV}\\
        m_{\rho}(N_c)&=(0.78\pm 0.05)-(0.02\pm 0.27)\frac{1}{N_c}\;\text{GeV},\\
        \Gamma_{\rho}(N_c)&=(0.48\pm 0.04)\frac{1}{N_c}-(0.08\pm 0.24)\frac{1}{N_c^2}\;\text{GeV}\\
         m_{f_2}(N_c)&=(1.16\pm 0.05)-(0.20\pm 0.31)\frac{1}{N_c}\;\text{GeV},\\
        \Gamma_{f_2}(N_c)&=(1.16\pm 0.10)\frac{1}{N_c}-(1.52\pm 0.52)\frac{1}{N_c^2}\;\text{GeV} \label{summary_resonances_last}
\end{align}
To view these results in another way, in fig. \ref{poles_Nc_s}, we show the motion of the $\sigma$, $\rho,f_2$ poles in the complex plane as $N_c$ is changed. We also compare with previous results obtained using unitarized $\chiPT$ for the $\sigma$ and $\rho$ poles \cite{Pelaez:2015qba}. 

\begin{figure}[h]
	\centering
    \includegraphics[width=\textwidth]{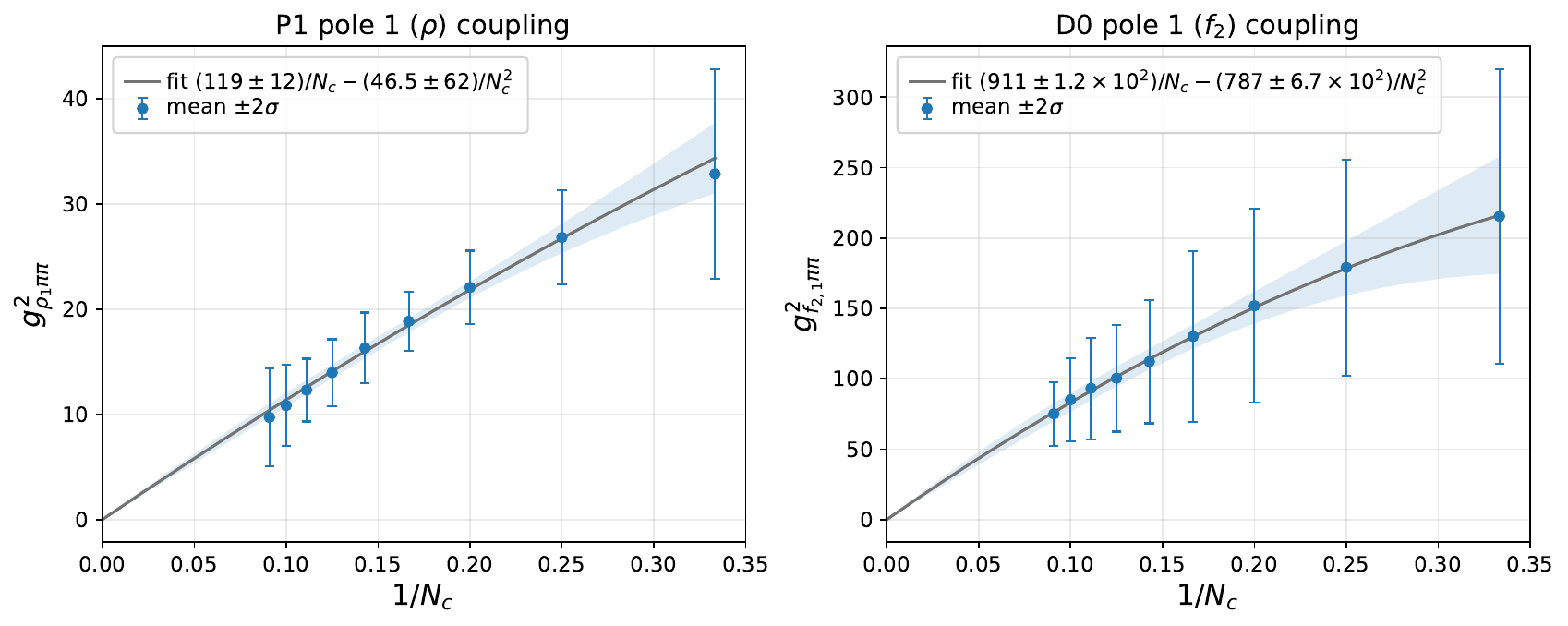}
	\caption{The couplings of the first $\rho,f_2$ resonances in the $P1,D0$ channels as function of $N_c$. The dots are the averaged results with $2\sigma$ error bars. The continuous plot are the fitted function of $1/N_c$ with a $1\sigma$ band. }
    \label{couplings}
\end{figure}

Finally, given the mass and width of the $\rho$ and $f_2$ resonances, we can extract the coupling of those resonances to the pions using the well-known expressions \cite{TheBook,EWERZ201431}:
\begin{equation}
\begin{aligned}
      \Gamma_\rho&=g_{\rho\pi\pi}^2\frac{m_\rho}{48\pi}\left[1-\frac{4m_\pi^2}{m_\rho^2}\right]^{\frac{3}{2}}\\
      \Gamma(f_2\rightarrow\pi^0\pi^0) &= \half\Gamma(f_2\rightarrow \pi^+\pi^-) = \frac{m_{f_2}}{960\pi} |g_{f_2\pi\pi}|^2 \left(\frac{m_{f_2}}{M_0}\right)^2 \left(1-\frac{4m_\pi^2}{m_{f_2}^2}\right)^\frac{5}{2}
\end{aligned}
\end{equation}
In fig. \ref{couplings}, we show the couplings of the $\rho$ and $f_2$ resonances as functions of $N_c$ and we extract the $N_c$ dependence of the couplings:
\begin{equation}
\begin{aligned}
        g^2_{\rho\pi\pi}&=(119\pm 12)\frac{1}{N_c}-(46.5\pm 62)\frac{1}{N_c^2},\\
        g^2_{f_2\pi\pi}&=(911\pm 120)\frac{1}{N_c}-(787\pm 670)\frac{1}{N_c^2}
\end{aligned}
\end{equation}

\subsection{Higher resonances}

\begin{figure}[h]
	\centering
    \includegraphics[width=\textwidth]{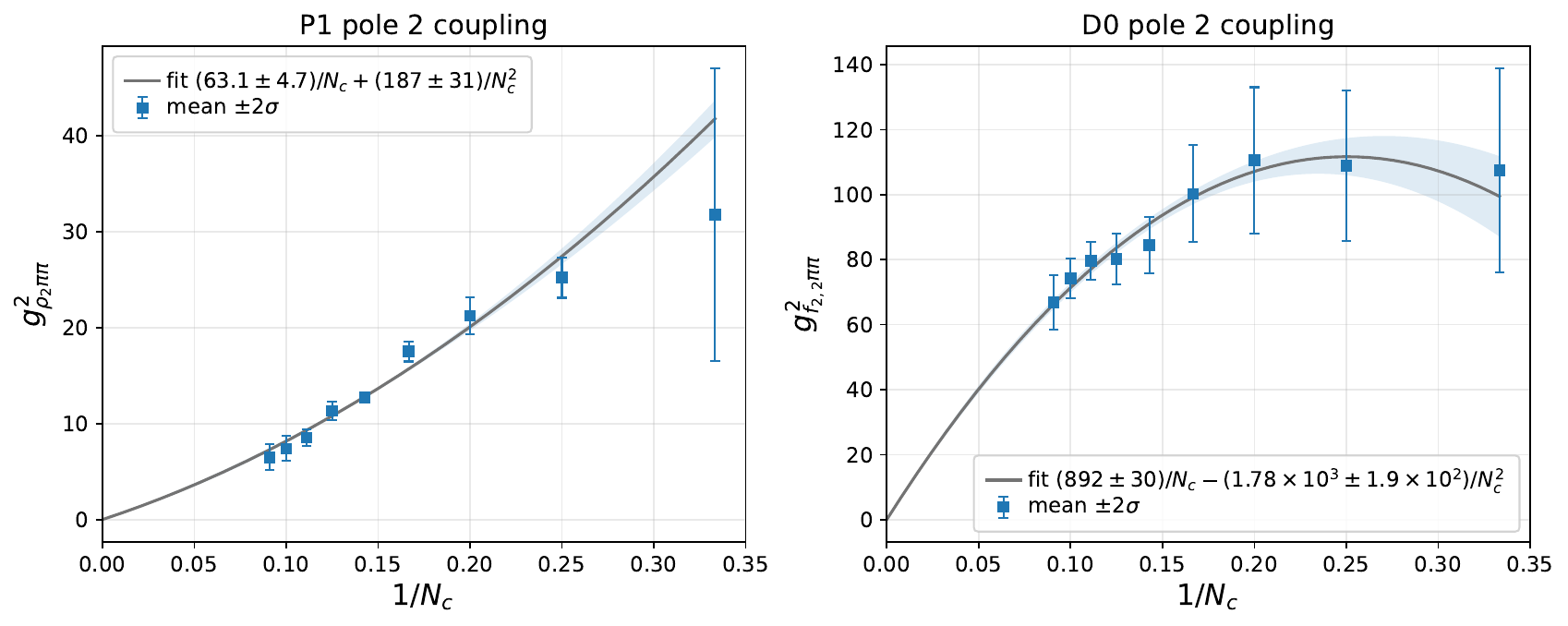}
	\caption{The couplings of the second $\rho$, $f_2$ resonances in the $P1$, $D0$ channels as function of $N_c^{-1}$. The discrete plots are the averaged results with $2\sigma$ error bars. The continuous plot are the fitted function of $1/N_c$ with a $1\sigma$ band.}
    \label{couplings2}
\end{figure}

\begin{figure}[t]
	\centering
	\includegraphics[width=\textwidth]{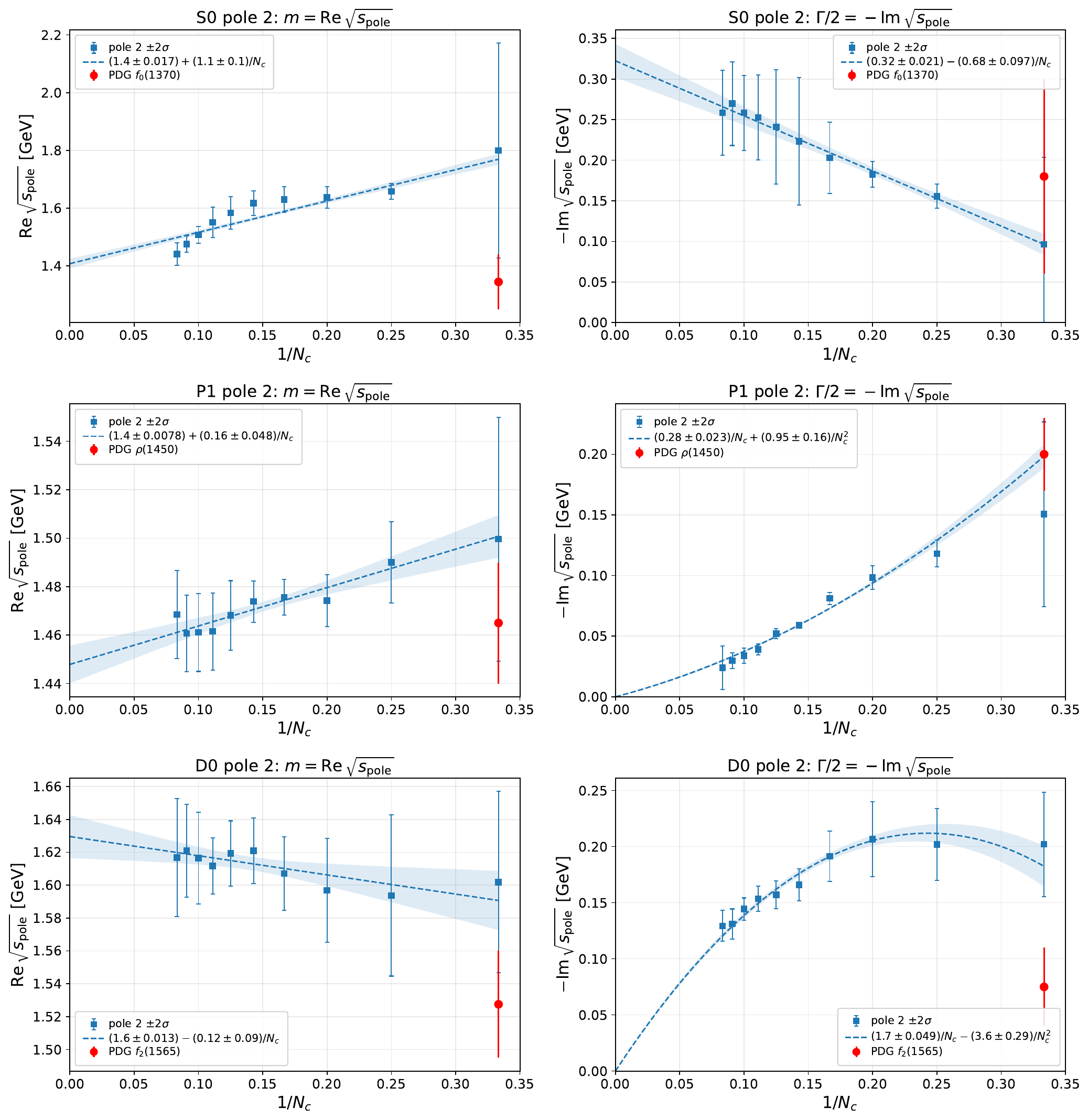}
	\caption{Real and imaginary parts of $\sqrt{s_\text{pole}}$ for the second $S0$, $P1$, $D0$ resonances as a function of $N_c^{-1}$. We present averaged results with $2\sigma$ error bars and the corresponding fitting function of $1/N_c$ with a $1\sigma$ band.}
	\label{poles_2_Nc}
\end{figure}

After studying the $\sigma$, $\rho$ and $f_2$ mesons we look at the second resonances in those channels. In the $P1$ and $D0$ channels those can be physically interpreted as excited states of the $\rho$ and $f_2$ mesons with the same quantum numbers. The real and imaginary parts of the poles are presented in fig. \ref{poles_2_Nc}. Their couplings to the pion are displayed in fig. \ref{couplings2}. 

Overall, these excited states do not depend as smoothly on $N_c$ as the leading $\sigma$, $\rho$, $f_2$ mesons which is likely due to low numerical precisions we have at higher energies. Nevertheless, in the $P1,D0$ channels, the masses and width of the second $\rho$ and $f_2$ poles follow the expected behavior of a $q\bar{q}$ state, with relatively stable mass and descreasing width with increasing $N_c$. The situation of the second pole in $S0$ channel is less clear. For $N_c=3$ comparing with experimental data ($N_f=2+1$), this state is above the $K\bar{K}$ production threshold and it is unclear how the analytic structure may be modified in the case of $N_f=2$ theory that we study here. We show the fit of the mass and width of this second $f_0$ state in fig. \ref{poles_2_Nc} for future reference.
 % In fact, continuing looking at the complex plane one can find more zeros but to have accurate extractions of their properties would require increasing the numerical resolutions at higher energies which we leave for future work.

\clearpage

\section{Conclusions}

The Gauge Theory Bootstrap framework provides a way to connect the low energy physics of asymptotically free gauge theories like QCD directly with their high energy (microscopic) descriptions and compute the strongly coupled dynamics in the intermediate energies. In this work, we turn on the $N_c$ dial of the GTB machinery and study gauge theories of different number of colors $N_c$. We obtain the $N_c$ dependence of the low energy parameters and of the meson resonances for a range of $N_c$.  The results compare well with alternative methods like unitarized $\chiPT$ \cite{Pelaez:2010er} and lattice gauge theory at finite and infinite $N_c$ \cite{DeGrand:2016pur,DeGrand:2023hzz,bonanno2025nonperturbativedeterminationmesonmasses}. It also agrees with the expected asymptotic behavior following from the $1/N_c$ expansion of QCD. For many quantities we evaluated not only the leading behavior as $N_c$ grows but also the subleading terms providing predictions for future calculations using other means. We also studied more systematically the sources of uncertainty due both to errors in the input parameters, mainly due to the asymptotic form factors, and to systematic errors due to variations in the implementation of the method. The consistency of the results with previous calculations when available and with general scaling expectations from large-$N_c$ provide further validation of the overall procedure.

\section*{Acknowledgement}

We are grateful to Fernando Romero-Lopez, Max Hansen and Claudio Bonanno for useful discussions.
Some of the results of this paper followed from computations performed on the Negishi community cluster, provided by the Rosen Center for Advanced Computing, Purdue University.

\newpage

\appendix
\counterwithin{equation}{section}

\section{Low energy formulas}\label{LEformula}

In this appendix we collect some additional formulas useful for the analysis of the low energy region. 

The scattering length are usually computed from the behavior of the phase shift (or partial wave) in the physical region near threshold as we did in the main text. However they can also be derived from the expansion of the amplitude near the symmetric point in the unphysical region (as in the Weinberg model). The resulting formulas are: 
\begin{subequations}\label{abrecompute}
\beqa
a^0_0 &=& 4\pi  \left(5 c_{00} + 8 c_{02} + \frac{16}{3} c_{10} - \frac{32}{3} c_{12} + \frac{224}{9} c_{20} + \frac{1408}{27} c_{30}+\ldots\right) \\
a^1_1 &=& \frac{16 \pi}{3}  \left(2c_{02} + c_{10} + \frac{4}{3} c_{12} - \frac{8}{3} c_{20} + \frac{16}{3}c_{30}+\ldots\right) \\
a^0_2 &=& \frac{16 \pi}{15}  \left(7 c_{02} + \frac{68}{3} c_{12} + 4 c_{20} - 16 c_{30}+\ldots \right)\\
a^2_0 &=& 8 \pi  \left(c_{00} + 4 c_{02} - \frac{4}{3} c_{10} - \frac{16}{3} c_{12} + \frac{16}{9} c_{20} - \frac{64}{27} c_{30}+\ldots \right) \\
b^0_0 &=& \frac{32 \pi}{3} \left(9 c_{02} + 3 c_{10} - 18 c_{12} + 28 c_{20} + 88 c_{30}+\ldots \right)\\
b^1_1 &=& \frac{16 \pi}{3}  \left(3 c_{02} + 4 c_{12} - 4 c_{20} + 16 c_{30}+\ldots\right)\\
b^0_2 &=& \frac{32\pi}{15} \left(17 c_{12} - 12 c_{30} +\ldots \right) \\
b^2_0 &=& 16 \pi  \left(6 c_{02} -  c_{10} - 12 c_{12} + \frac{8}{3} c_{20} - \frac{16}{3} c_{30}  +\ldots \right)
\eeqa
\end{subequations}
that can be used to check the physical region calculation. 

\section{High energy 2-point functions}\label{higherloops}

The two point functions in \eqref{PIS}, \eqref{PIV}, as well as the one for \eqref{jq} can be computed to higher order. From \cite{Chetyrkin:1996sr,Chetyrkin:1996ez,Gorishnii:1990vf,Zoller:2012qv,Bagan:1988ay,narison_2004} we obtain:
\begin{equation}\label{2pt}
	\begin{aligned}
		\Pi^0_0(s) = &-\frac{s}{(2\pi)^4}\frac{ m_q^2}{8\pi^2}  \Bigg\{ 2N_c \ln\left(-\frac{s}{\mu^2}\right) +\frac{\alpha_s}{\pi} \left[ s_1\ln \left(-\frac{s}{\mu^2}\right)+s_2\ln^2\left(-\frac{s}{\mu^2}\right)\right]  \\
		&  +\left(\frac{\alpha_s}{\pi}\right)^2 
		\left[s_3\ln \left(-\frac{s}{\mu^2}\right)+s_4 \ln^2 \left(-\frac{s}{\mu^2}\right)+s_5\ln^3 \left(-\frac{s}{\mu^2}\right)\right]+O(\alpha_s^3)\Bigg\}\\
        \Pi^1_1(s) =  &-\frac{s}{(2\pi)^4}\frac{ N_c }{24\pi^2} \Bigg\{\ln\left(-\frac{s}{\mu^2}\right) +r_1\frac{\alpha_s}{\pi} \ln\left(-\frac{s}{\mu^2}\right)  \\
		&  +\left(\frac{\alpha_s}{\pi}\right)^2 
		\left[ r_2\ln \left(-\frac{s}{\mu^2}\right)+r_3 \ln^2 \left(-\frac{s}{\mu^2}\right)\right]+O(\alpha_s^3) \Bigg\}\\
        \Pi_2^0(s)=&\frac{2s^2}{(2\pi)^4}\frac{1}{16\pi^2}\Bigg\{-\frac{N_c}{10}\ln\left(-\frac{s}{\mu^2}\right)+\frac{\alpha_s}{\pi}\left(a_1\ln\left(-\frac{s}{\mu^2}\right)+a_2\ln^2\left(-\frac{s}{\mu^2}\right)\right)+O(\alpha_s^2)\Bigg\}
	\end{aligned}
\end{equation}
with the coefficients:
\begin{equation}
	\begin{aligned}
		s_1&=\frac{17 \left(N_c^2-1\right)}{4},\;\; s_2=-\frac{3 \left(N_c^2-1\right)}{4}\\
		s_3&=-\frac{\left(N_c^2-1\right) \left(N_c^2 (640 \zeta (3)-2477)-8 N_c (16 \zeta (3)-65)-144 \zeta (3)+691\right)}{128 N_c}\\
		s_4&=-\frac{\left(N_c^2-1\right) \left(883 N_c^2-176 N_c-315\right)}{192 N_c},\;\; s_5=\frac{\left(N_c^2-1\right) \left(20 N_c^2-4 N_c-9\right)}{48 N_c}\\
        r_1&=\frac{3 \left(N_c^2-1\right)}{8 N_c},\;\; r_3=-\frac{(11 N_c-4) \left(N_c^2-1\right)}{64 N_c}\\
		r_2&=-\frac{\left(N_c^2-1\right) \left(N_c^2 (176 \zeta (3)-243)+N_c (88-64 \zeta (3))-3\right)}{128 N_c^2}\\
        a_1&=\frac{N_c^2-1}{10}\frac{473}{360},\;\;a_2=-\frac{N_c^2-1}{30}
	\end{aligned}
\end{equation}
These expressions are the ones that we use in the actual numerical calculations. 

\section{Numerical implementation}

The numerical implementation is similar to the previous work on GTB where the set of bootstrap variables are discretized and constraints of unitarity, QCD sum rules, asymptotic form factors and low energy matching are imposed in respective kinematic regions. In this work, we discretized the bootstrap variables using the discrete basis previously employed in \cite{He:2021eqn} for the scattering amplitude and extended an implemented in \cite{Cordoba:2025wnv} for GTB in the $N_c=3$ case. As advocated in that reference, it has the advantage of speeding up the numerical optimization procedure, as well as facilitating the extraction of the resonance poles on the complex plane of $s$. In this appendix, we briefly describe the numerical implementation. Since the GTB program can be found in the github repository \href{https://github.com/hyfysics/gauge-theory-bootstrap}{gauge-theory-bootstrap}, the appendix mainly serves as a guide for interested reader to work through the numerical program.

\medskip

As is standard in the S-matrix bootstrap \cite{Paulos:2017fhb}, cut-planes of kinematic variable $\nu$ (that stands for $s,t,u$) are mapped to unit disk through
\beq
z = \frac{\sqrt{4-\nu_0}-\sqrt{4-\nu}}{\sqrt{4-\nu_0}+\sqrt{4-\nu}}, \ \ \ \ z(x+i\epsilon)=e^{i\varphi}, \ \ \ \ x\in \mathbb{R}_{>4}
\label{a161}
\eeq
where the physical regions above the cut is mapped to the upper half circle and the physical energy $x$ is related to the angle $\varphi$ through
\beq
x= \nu_0+\frac{8-2\nu_0}{1+\cos\frac{\varphi}{2}} 
\label{a159}
\eeq
The amplitude parameters $\sigma,\rho$ are then expanded by the basis functions $\phi_n=\sin(n\varphi)$ as \cite{He:2021eqn}
\begin{equation}
 	\sigma(x) = \sum_{n=1}^\infty \sigma_n \phi_n(x),\;\; \rho(x,y) = \sum_{n,m=1}^\infty \rho_{nm} \phi_{n}(x) \phi_{m}(y)  
\end{equation}
Defining
\begin{subequations}
	\beqa
    \Phi_{n}(\nu) &=& \frac{1}{\pi} \int_4^\infty\!\!\!\!\!\! dx\ \frac{\phi_{n}(x)}{x-\nu}= z^n\\
	\hat{\Phi}_{n,\ell}(s) &=& \int_{-1}^{+1} d\mu P_\ell(\mu) \Phi_{n}(t)  \label{a164}\\
	\tilde{\Phi}_{nm,\ell}(s) &=& \int_{-1}^{+1} d\mu  P_\ell(\mu) \Phi_{n}(t)\Phi_{m}(u)  \label{a165}
	\eeqa
\end{subequations}
the amplitude $A(s,t,u)$ and the partial waves $h^I_\ell(s)$ are expanded as
\beq\label{Astu}
\begin{aligned}
A(s,t,u) = &f_0 + \sum_{n=1}^\infty \left[\sigma^{(1)}_n \Phi_n(s) + \sigma^{(2)}_n (\Phi_n(t)+\Phi_n(u))\right]\\
+&\sum_{n,m=1}^\infty \left[\rho^{(1)}_{nm} \Phi_n(s) (\Phi_m(t)+\Phi_m(u)) + \rho^{(2)}_{nm} \Phi_n(t)\Phi_m(u) \right]
\end{aligned}
\eeq
and
\beq
\begin{aligned}
h^I_\ell(s) = &\frac{\pi}{4} \sqrt{\frac{s-4}{s}} \Biggl\{f_0 A^I_\ell + \sum_{n=1}^\infty \left[ A^I_{\ell n}(s)\sigma^{(1)}_n + B^I_{\ell n}(s) \sigma^{(2)}_n \right] \\
+ &\sum_{n,m=1}^\infty  \left[ A^I_{\ell,nm}(s)\ \rho^{(1)}_{nm} + B^I_{\ell,nm}(s)\ \rho^{(2)}_{nm}\right] \biggr\} \ ,  
\end{aligned}\label{a176}
\eeq
with
\beq
\begin{aligned}
	A^0_\ell &= 10\delta_{\ell 0},\;   A^0_{\ell,n} = 6\delta_{\ell 0}\Phi_n+ 2\hat{\Phi}_{n\ell},\; B^0_{\ell,n} = 4\delta_{\ell 0}\Phi_n+ 8\hat{\Phi}_{n\ell},\\
    A^0_{\ell,nm} &= 6\Phi_n\hat{\Phi}_{m\ell}+ 2\Phi_m\hat{\Phi}_{n\ell}+2\tilde{\Phi}_{nm,\ell},\;
    B^0_{\ell,nm} =  \Phi_n\hat{\Phi}_{m\ell}+\Phi_m\hat{\Phi}_{n\ell}+ 3\tilde{\Phi}_{nm,\ell} , \\
    A^1_\ell &= 0,\;  A^1_{\ell,n}= 2 \hat{\Phi}_{n\ell}, \;
    B^1_{\ell,n} =  -2\hat{\Phi}_{n\ell},\\
    A^1_{\ell,nm}&= 2 \Phi_m\hat{\Phi}_{n\ell}+2\tilde{\Phi}_{nm,\ell},\;
    B^1_{\ell,nm} =  -\Phi_n\hat{\Phi}_{m\ell}-\Phi_m\hat{\Phi}_{n\ell},\\
    A^2_\ell &= 4\delta_{\ell 0}, \;	 A^2_{\ell,n} = 2 \hat{\Phi}_{n\ell}, \;   B^2_{\ell,n} = 4 \delta_{\ell 0} \Phi_n + 2 \hat{\Phi}_{n\ell}, \\
	A^2_{\ell,nm} &= 2 \Phi_m\hat{\Phi}_{n\ell}+2\tilde{\Phi}_{nm,\ell}    \ ,
	B^2_{\ell,nm} =  \Phi_n\hat{\Phi}_{m\ell}+\Phi_m\hat{\Phi}_{n\ell}.
\end{aligned}
\eeq
With the set of variables $\{f_0,\sigma_n^{(1,2)},\rho_{nm}^{(1,2)}\}$, one can then access the full analytic functions of $A(s,t,u)$ and $h^I_\ell(s)$ in any kinematic regions.

The form factors can be analogously expanded as series on the unit disk:
\begin{equation}
	F^{I}_{\ell}(z)=\sum_{n=1}^\infty b^{I}_{\ell,n}z^n
\end{equation}
and the normalization condition
\begin{equation}
	F^I_{\ell}\big(z(0)\big)=F^{I}_{\ell,0}
\end{equation}
is a set of linear constraints on the $b^I_\ell$.
The real and imaginary parts in the physical region are evaluated on the upper half circle as
\begin{equation}\label{RIF}
	\text{Re}F^{I}_{\ell}(\varphi)=\sum_{n=1}^{\infty} b^{I}_{\ell,n}\cos(n\varphi),\;\;\text{Im}F^{I}_{\ell}(\varphi)=\sum_{n=1}^{\infty} b^{I}_{\ell,n}\sin(n\varphi)
\end{equation}

For parameterizing the spectral density, in \cite{He:2025gws}, it was shown that a parameterization taking into account the low energy free pion kinematics \eqref{lowerho} greatly reduced the parameter space. Here we slightly improve it by incorporating the high energy behavior of the spectral density with $N_c$. In particular, we parameterized the rescaled spectral density $\hat{\rho}^I_\ell(s)$ defined through:
\begin{equation}
\rho^I_\ell=\kappa_\ell(s)\hat{\rho}^I_\ell(s)
\end{equation}
where $\kappa_\ell$ are the free kinematic factors equivalently the expressions in \eqref{lowerho}. Note that this rescaled spectral density enters directly into the modified positive semidefinite matrix \eqref{sdpdiscrete} below. At high energies, the $\hat{\rho}^I_\ell$ goes as
\begin{equation}
    \begin{aligned}
        \hat{\rho}^0_0(s\to \infty)&=\frac{8m_q^2N_c s}{3}\\
        \hat{\rho}^1_1(s\to \infty)&=2N_c\\
        \hat{\rho}^0_2(s\to \infty)&=4N_c
    \end{aligned}
\end{equation}
and by definition, they approach 1 near threshold.
We then write the $\hat{\rho}^I_\ell$ using a basis that includes an interpolation function between the UV and IR limit:\footnote{The form for $\psi^0_2(s)$ takes into account the form factor normalization \eqref{Fq0} and the expected saturation $|F|^2=\hat{\rho}$ at low energy.}
\begin{equation}
\begin{aligned}
    \psi^0_0(s)&=\frac{8m_q^2N_c s}{3}+e^{-s/100}\\
    \psi^1_1(s)&=2N_c+e^{-s/100}(1-2N_c)\\
    \psi^0_2(s)&=\frac{N_f^2}{(4C_F+N_f)^2}+e^{-s/100}\left(1-\frac{N_f^2}{(4C_F+N_f)^2}\right)
\end{aligned}
\end{equation}
plus a set of functions suppressed near threshold ($\varphi(s)$ here is the inverse function of \eqref{a159}):
\begin{equation}
    \tilde{\psi}^I_{\ell,n}=\big(1-4/s\big)\sin(n\varphi(s))
\end{equation}
and the spectral densities $\hat{\rho}^I_\ell$ are expanded as
\begin{equation}
\hat{\rho}^I_\ell=a_0\psi^I_\ell+\sum_{n=1}a_n\tilde{\psi}^I_{\ell,n}
\end{equation}
This gives an example of one possible parameterization of the spectral density that takes into account the UV and IR behavior. Several other similar choices give similar results. It is an interesting open question to find a complete set of basis that encodes automatically the UV and IR behaviors of the spectral densities.

In the numeric computations, all the sums $\sum_{n=1}^\infty$ in the above parameterizations are truncated to certain finite order. The convergence of such truncation order was tested in \cite{Cordoba:2025wnv}.

\smallskip

The physical region $s>4m_\pi^2$ is discretized at a set of points equivalent to equally spaced points on the upper half circle:
\begin{equation}\label{xi}
\begin{aligned}
    s_i&=\nu_0+\frac{8-2\nu_0}{1+\cos\frac{\varphi_i}{2}}\\
    \varphi_i&=\left(j-\frac{1}{2}\right)\Delta_{\phi},\;\;j=1,2,\ldots, M,\;\;\Delta_{\phi}=\frac{\pi}{M}
\end{aligned}
\end{equation}
and the positive semidifnite condition \ref{Bmt} for $S0, P1, D0$ waves is imposed at this set of energies up to the energy scale $s_0$ beyond which perturbative QCD should be valid. In practice, we impose the positive semidefinite conditions \cite{He:2024nwd}:\footnote{This comes from constructing the matrix from overlap of linear combinations of the original three states. See section 8.4 or \cite{He:2024nwd} for more details.}
\begin{equation}\label{sdpdiscrete}
	\begin{aligned}
		\begin{pmatrix}
			1&\tilde{h}^I_{\ell}(s_i)&F^I_{\ell}(s_i)\\
			\tilde{h}^{*I}_{\ell}(s_i)&2\text{Im}\hat{h}^I_{\ell}(s_i)&2\text{Im}\hat{F}^I_{\ell}(s_i)\\
			F^{*I}_{\ell}(s_i)&2\text{Im}\hat{F}^I_{\ell}(s_i)&\hat{\rho}^I_{\ell}(s_i)
		\end{pmatrix}\succeq 0,\;s_i=1,\ldots,n_0,\;\;\{\ell,I\}=S0,P1,D0
	\end{aligned}
\end{equation}
where
\begin{equation}
    \begin{aligned}
    \tilde{h}^I_\ell(s)=\Lambda_\ell^{-1}(s)h^I_\ell(s),\;\;\text{Im}\hat{h}^I_\ell(s)=\Lambda^{-2}_\ell(s)h^I_\ell(s)\\
\text{Im}\hat{F}^I_\ell(s)=\Lambda^{-1}_\ell(s)\text{Im}F^I_\ell(s)
    \end{aligned}
\end{equation}
with
\begin{equation}
\Lambda_\ell(s)=\left(\frac{\sqrt{s}-2}{\sqrt{s}+2}\right)^{\frac{\ell}{2}}
\end{equation}

For other quantum numbers, the simple partial-wave unitarity \eqref{unit} is imposed at all physical energies:
\begin{equation}
	(\text{Re}\tilde{h}^{I}_\ell)^2+(\text{Im}\tilde{h}^{I}_\ell)^2\le 2\text{Im}\hat{h}^{I}_\ell,\;\; s_i=1,\ldots, M,\;\; \forall I,\ell.
\end{equation}

To input the UV information from the perturbative computations, a scale $s_0=2\text{GeV}$ is chosen where
\begin{equation}
	\alpha_s(s_0)\simeq 0.315,\;\; m_q(s_0)\simeq 5.25\text{MeV}
\end{equation}
In the numerical computations, all the dimensionful quantities are expressed using the pion mass $m_\pi=140\text{ MeV}$ as mass unit.
For the scalar, vector and tensor currents $j_S,j_V,j_T$, the sum rules are imposed for $n_{SR}$ moments to agree with the perturbative QCD values up to a small error:\footnote{Here $l=1$ for $j_{S,V}$ and $l=2$ for $j_T$.}
\begin{equation}
	\vline\frac{1}{s_0^{n+l+1}}\int_4^{s_0}dx\rho(x)w_n(x)-\frac{1}{s_0^{n+l+1}}\int_4^{s_0}dx\rho_{pQCD}(x)w_n(x)\vline\le \epsilon^I_{l,n}
\end{equation}
where $w_n(x)$ is a series of weight functions. \footnote{We use the ``pinch-weight" concentrating on the low energy region that was previously used in \cite{cherry2001qcd} for phenomenological studies. See e.g. appendix C of \cite{He:2025gws}.} In the numerics, we choose the sum rule errors $\epsilon^I_{\ell,n}$ to be a small percentage of the sum rules themselves. This takes care of how the sum rule errors may vary with the $N_c$ parameter. Furthermore, in the region with energy above $s_0$, we require that the spectral densities to agree with that of the perturbative QCD values. In practice, we constrain
\begin{equation}
    |\rho^I_\ell(s_i)-\rho^I_{\ell,\text{pQCD}}(s_i)|\le \varepsilon(s),\; i=n_0+1,\ldots,M
\end{equation}
where $\varepsilon(s)$ are functions that descrease with $s$. Several different choices of $\varepsilon(s)$ were tested in the numerical program with similar low energy results. One choice can be found in the numerical program in the repository. 

To constrain the asymptotic form factor, we implement the constraints \eqref{chiIL} at a discrete set of points above $s_0$ with $i=n_0+1,\ldots,M$ in \eqref{xi}:
\begin{subequations}\label{asympF}
	\beqa
	|F_0^0(s_i>s_0)| & \le & \chi_0^0\ \frac{m_{\pi}^4}{s}\, \ln\!\left(\frac{s}{m_{\pi}^2}\right) \label{scalarasymFF}\\
	|F_1^1(s_i>s_0)| & \le  & \chi_1^1\ \frac{16\pi\alpha_s f_\pi^2}{s}\\
	|F_2^0(s_i>s_0)| & \le  & \chi_2^0\ \frac{48\pi\alpha_s f_\pi^2}{s}
	\eeqa
\end{subequations}
As described in section \ref{err}, to account for this unknown numerical factor, we take a set of values \eqref{chirange} and average to obtain mean values of each result and an estimate of error.

Finally, as described in \eqref{chimatch}, the partial amplitudes are matched with the tree level computation in the unphysical region $s\le2m^2_{\pi}$ at a discrete set of points as in \eqref{chimatch} up to some tolerance $\epsilon^{\chi}$ which we change with $N_c$ conservatively as \eqref{echiNc}.

With the above constraints, the computation then proceeds with the Watsonian iterative procedure as described in section \ref{GTBreview}. To obtain the results shown in this paper, we use two alternative procedures. In one, we run each $N_c$ independently following what was done for $N_c=3$ in previous work. In this procedure, when $N_c\ge 6$, the single $f_2$ peak in the $D0$ channel contains two near-by poles instead of a single $f_2$ pole. One way to cure this problem is to start the iterations with a solution that already has a single pole in the D0 channel form factor (in this case we choose $s^{\text{D0}}_{\text{seed-pole}}\simeq 72.45- 5.2 i$ in $\mpi=1$ units, which is the position obtained in the initial run for the largest value of $N_c$). The pole\footnote{Notice that these poles are on the second sheet, so there is never an actual pole in the physical sheet. What actually changes with the iterations is the shape of the resonance peak since, in the iterations, the parameterizations do not have poles.} moves as we iterate and eventually finds its natural position for the corresponding value of $N_c$. For the $S0$ and $P1$ channels the calculation is unmodified although one can think of including 'seed-poles' in those channels too. Alternatively, in the other procedure we use, the converged results of a given $N_c$ are used in the initial functional to start the iteration procedure for the next $N_c$ value. In this case the $N_c$ solution contains all the resonances in the various channels and the subsequent iteration moves the resonances to their natural positions for the next $N_c$ value. In that way we obtain the solutions that can be considered the natural extension of the $N_c=3$ solution to higher values of $N_c$. To be more concrete, in this last case, for the initial functionals to maximize for starting the iterative procedure of the $N_c+1,N_c\ge 3$ computation, we maximize the functional
\begin{equation}
    \begin{aligned}
    \text{max}\;&\mathcal{F}_1+\mathcal{F}_2,\;\; \\
       \mathcal{F}_1=&\sum_{s_i=1}^{n_0}\sum_{I,\ell=S0,P1,D0}\left\{\frac{|h^I_\ell|}{|F^I_\ell|}\bigg|_{N_c}(\text{Re}F^I_{\ell}|_{N_c}\text{Re}h^I_{\ell}|_{\text{new}}+\text{Im}F^I_{\ell}|_{N_c}\text{Im}h^I_{\ell}|_{\text{new}})-\text{Im}h^I_{\ell}|_{\text{new}}\right\}\\
    \mathcal{F}_2=&\sum_{s_i=1}^{n_0}\sum_{I,\ell=S0,P1,D0}\left\{2\text{Re}\mathcal{F}^I_{\ell}|_{N_c}\text{Re}\mathcal{F}^I_{\ell}|_{\text{new}}+2\text{Im}\mathcal{F}^I_{\ell}|_{N_c}\text{Im}\mathcal{F}^I_{\ell}|_{\text{new}}-\rho^I_\ell|_{\text{new}}\right\}
        \end{aligned}
\end{equation}
where subscript $_{N_c}$ denotes the converged results from the previous number of flavor. In the case of computing $N_c=3$, the initial functional is simply $\text{max}\;0$. After this initial functional, the usual iterations proceeds until the results converges before the next value of $N_c$. Typically it takes 10-20 iterations for the computations for each $N_c$ to converge.

\bibliographystyle{utphys}
\bibliography{gtb.bib}

\providecommand{\href}[2]{#2}\begingroup\raggedright\begin{thebibliography}{10}

\bibitem{PhysRevLett.133.191601}
Y.~He and M.~Kruczenski, ``Gauge theory bootstrap,''
  \href{http://dx.doi.org/10.1103/PhysRevLett.133.191601}{{\em Phys. Rev.
  Lett.} {\bfseries 133} (Nov, 2024) 191601}.
  \url{https://link.aps.org/doi/10.1103/PhysRevLett.133.191601}.

\bibitem{PhysRevD.110.096001}
Y.~He and M.~Kruczenski, ``Bootstrapping gauge theories,''
  \href{http://dx.doi.org/10.1103/PhysRevD.110.096001}{{\em Phys. Rev. D}
  {\bfseries 110} (Nov, 2024) 096001}.
  \url{https://link.aps.org/doi/10.1103/PhysRevD.110.096001}.

\bibitem{He:2024nwd}
Y.~He and M.~Kruczenski, ``{Gauge Theory Bootstrap: Pion amplitudes and low
  energy parameters},'' \href{http://arxiv.org/abs/2403.10772}{{\ttfamily
  arXiv:2403.10772 [hep-th]}}.

\bibitem{He:2025gws}
Y.~He and M.~Kruczenski, ``{The Gauge Theory Bootstrap: Predicting Pion
  Dynamics from QCD},'' \href{http://dx.doi.org/10.1093/ptep/ptag087}{{\em
  PTEP} {\bfseries 2026} no.~6, (2026) 063B05},
  \href{http://arxiv.org/abs/2505.19332}{{\ttfamily arXiv:2505.19332
  [hep-th]}}.

\bibitem{Paulos:2016but}
M.~F. Paulos, J.~Penedones, J.~Toledo, B.~C. van Rees, and P.~Vieira, ``{The
  S-matrix bootstrap II: two dimensional amplitudes},''
  \href{http://dx.doi.org/10.1007/JHEP11(2017)143}{{\em JHEP} {\bfseries 11}
  (2017) 143}, \href{http://arxiv.org/abs/1607.06110}{{\ttfamily
  arXiv:1607.06110 [hep-th]}}.

\bibitem{Paulos:2016fap}
M.~F. Paulos, J.~Penedones, J.~Toledo, B.~C. van Rees, and P.~Vieira, ``{The
  S-matrix bootstrap. Part I: QFT in AdS},''
  \href{http://dx.doi.org/10.1007/JHEP11(2017)133}{{\em JHEP} {\bfseries 11}
  (2017) 133}, \href{http://arxiv.org/abs/1607.06109}{{\ttfamily
  arXiv:1607.06109 [hep-th]}}.

\bibitem{Paulos:2017fhb}
M.~F. Paulos, J.~Penedones, J.~Toledo, B.~C. van Rees, and P.~Vieira, ``{The
  S-matrix bootstrap. Part III: higher dimensional amplitudes},''
  \href{http://dx.doi.org/10.1007/JHEP12(2019)040}{{\em JHEP} {\bfseries 12}
  (2019) 040}, \href{http://arxiv.org/abs/1708.06765}{{\ttfamily
  arXiv:1708.06765 [hep-th]}}.

\bibitem{Karateev:2019ymz}
D.~Karateev, S.~Kuhn, and J.~Penedones, ``{Bootstrapping Massive Quantum Field
  Theories},'' \href{http://dx.doi.org/10.1007/JHEP07(2020)035}{{\em JHEP}
  {\bfseries 07} (2020) 035}, \href{http://arxiv.org/abs/1912.08940}{{\ttfamily
  arXiv:1912.08940 [hep-th]}}.

\bibitem{Shifman:1978bw}
M.~A. Shifman, A.~I. Vainshtein, and V.~I. Zakharov, ``{QCD and Resonance
  Physics. The rho-omega Mixing},''
  \href{http://dx.doi.org/10.1016/0550-3213(79)90024-5}{{\em Nucl. Phys. B}
  {\bfseries 147} (1979) 519--534}.

\bibitem{SHIFMAN1979385}
M.~Shifman, A.~Vainshtein, and V.~Zakharov, ``Qcd and resonance physics.
  theoretical foundations,''
  \href{http://dx.doi.org/https://doi.org/10.1016/0550-3213(79)90022-1}{{\em
  Nuclear Physics B} {\bfseries 147} no.~5, (1979) 385--447}.

\bibitem{SHIFMAN1979448}
M.~Shifman, A.~Vainshtein, and V.~Zakharov, ``Qcd and resonance physics.
  applications,''
  \href{http://dx.doi.org/https://doi.org/10.1016/0550-3213(79)90023-3}{{\em
  Nuclear Physics B} {\bfseries 147} no.~5, (1979) 448--518}.

\bibitem{osti_1447331}
G.~Peter~Lepage and S.~J. Brodsky, ``Exclusive processes in quantum
  chromodynamics: Evolution equations for hadronic wavefunctions and the form
  factors of mesons,''
  \href{http://dx.doi.org/10.1016/0370-2693(79)90554-9}{{\em Physics Letters B}
  {\bfseries 87} no.~4, (11, 1979) }.

\bibitem{Lepage:1980fj}
G.~P. Lepage and S.~J. Brodsky, ``{Exclusive Processes in Perturbative Quantum
  Chromodynamics},'' \href{http://dx.doi.org/10.1103/PhysRevD.22.2157}{{\em
  Phys. Rev. D} {\bfseries 22} (1980) 2157}.

\bibitem{PhysRevLett.17.616}
S.~Weinberg, ``{Pion scattering lengths},''
  \href{http://dx.doi.org/10.1103/PhysRevLett.17.616}{{\em Phys. Rev. Lett.}
  {\bfseries 17} (1966) 616--621}.

\bibitem{GASSER1984142}
J.~Gasser and H.~Leutwyler, ``Chiral perturbation theory to one loop,''
  \href{http://dx.doi.org/https://doi.org/10.1016/0003-4916(84)90242-2}{{\em
  Annals of Physics} {\bfseries 158} no.~1, (1984) 142--210}.

\bibitem{Cordoba:2025wnv}
R.~Cordoba, ``{Discrete Basis Parameterization for the Gauge Theory
  Bootstrap},'' \href{http://arxiv.org/abs/2511.11513}{{\ttfamily
  arXiv:2511.11513 [hep-th]}}.

\bibitem{Zahed:2024haa}
A.~Zahed, ``{Bootstrapping the leading hadronic muon anomaly},''
  \href{http://dx.doi.org/10.1103/3gs1-8s82}{{\em Phys. Rev. D} {\bfseries 112}
  no.~3, (2025) 034031}, \href{http://arxiv.org/abs/2412.00187}{{\ttfamily
  arXiv:2412.00187 [hep-th]}}.

\bibitem{tHooft:1973alw}
G.~'t~Hooft, ``{A Planar Diagram Theory for Strong Interactions},''
  \href{http://dx.doi.org/10.1016/0550-3213(74)90154-0}{{\em Nucl. Phys. B}
  {\bfseries 72} (1974) 461}.

\bibitem{Witten:1979kh}
E.~Witten, ``{Baryons in the 1/n Expansion},''
  \href{http://dx.doi.org/10.1016/0550-3213(79)90232-3}{{\em Nucl. Phys. B}
  {\bfseries 160} (1979) 57--115}.

\bibitem{DeGrand:2016pur}
T.~DeGrand and Y.~Liu, ``{Lattice study of large $N_c$ QCD},''
  \href{http://dx.doi.org/10.1103/PhysRevD.94.034506}{{\em Phys. Rev. D}
  {\bfseries 94} no.~3, (2016) 034506},
  \href{http://arxiv.org/abs/1606.01277}{{\ttfamily arXiv:1606.01277
  [hep-lat]}}. [Erratum: Phys.Rev.D 95, 019902 (2017)].

\bibitem{Hernandez:2019qed}
P.~Hern{\'a}ndez, C.~Pena, and F.~Romero-L{\'o}pez, ``{Large $N_c$ scaling of
  meson masses and decay constants},''
  \href{http://dx.doi.org/10.1140/epjc/s10052-019-7395-y}{{\em Eur. Phys. J. C}
  {\bfseries 79} no.~10, (2019) 865},
  \href{http://arxiv.org/abs/1907.11511}{{\ttfamily arXiv:1907.11511
  [hep-lat]}}.

\bibitem{Baeza-Ballesteros:2022azb}
J.~Baeza-Ballesteros, P.~Hern{\'a}ndez, and F.~Romero-L{\'o}pez, ``{A lattice
  study of {\ensuremath{\pi}}{\ensuremath{\pi}} scattering at large N$_{c}$},''
  \href{http://dx.doi.org/10.1007/JHEP06(2022)049}{{\em JHEP} {\bfseries 06}
  (2022) 049}, \href{http://arxiv.org/abs/2202.02291}{{\ttfamily
  arXiv:2202.02291 [hep-lat]}}.

\bibitem{DeGrand:2023hzz}
T.~A. DeGrand and E.~Wickenden, ``{Lattice study of the chiral properties of
  large-Nc QCD},'' \href{http://dx.doi.org/10.1103/PhysRevD.108.094516}{{\em
  Phys. Rev. D} {\bfseries 108} no.~9, (2023) 094516},
  \href{http://arxiv.org/abs/2309.12270}{{\ttfamily arXiv:2309.12270
  [hep-lat]}}.

\bibitem{DeGrand:2024lvp}
T.~DeGrand, ``{Curve collapse for the isospin-2 pion scattering length from QCD
  with 3, 4, and 5 colors},'' \href{http://arxiv.org/abs/2409.02242}{{\ttfamily
  arXiv:2409.02242 [hep-lat]}}.

\bibitem{Baeza-Ballesteros:2025iee}
J.~Baeza-Ballesteros, P.~Hern{\'a}ndez, and F.~Romero-L{\'o}pez, ``{The
  {\ensuremath{\pi}}{\ensuremath{\pi}} scattering amplitude at large
  N$_{c}$},'' \href{http://dx.doi.org/10.1007/JHEP08(2025)110}{{\em JHEP}
  {\bfseries 08} (2025) 110}, \href{http://arxiv.org/abs/2503.13978}{{\ttfamily
  arXiv:2503.13978 [hep-lat]}}.

\bibitem{bonanno2025nonperturbativedeterminationmesonmasses}
C.~Bonanno, M.~G. Pérez, A.~González-Arroyo, K.-I. Ishikawa, and M.~Okawa,
  ``Non-perturbative determination of meson masses and low-energy constants in
  large-$n$ qcd,'' 2025.
\newblock \url{https://arxiv.org/abs/2508.05446}.

\bibitem{Pelaez:2006nj}
J.~R. Pelaez and G.~Rios, ``{Nature of the f0(600) from its N(c) dependence at
  two loops in unitarized Chiral Perturbation Theory},''
  \href{http://dx.doi.org/10.1103/PhysRevLett.97.242002}{{\em Phys. Rev. Lett.}
  {\bfseries 97} (2006) 242002},
  \href{http://arxiv.org/abs/hep-ph/0610397}{{\ttfamily arXiv:hep-ph/0610397}}.

\bibitem{Albert:2026xyz}
J.~Albert, D.~Kosva, and L.~Rastelli, ``{Bootstrapping Pion Form Factors at
  Large $N$},'' \href{http://arxiv.org/abs/2606.19420}{{\ttfamily
  arXiv:2606.19420 [hep-th]}}.

\bibitem{Guerrieri:2018uew}
A.~L. Guerrieri, J.~Penedones, and P.~Vieira, ``{Bootstrapping QCD Using Pion
  Scattering Amplitudes},''
  \href{http://dx.doi.org/10.1103/PhysRevLett.122.241604}{{\em Phys. Rev.
  Lett.} {\bfseries 122} no.~24, (2019) 241604},
  \href{http://arxiv.org/abs/1810.12849}{{\ttfamily arXiv:1810.12849
  [hep-th]}}.

\bibitem{Bose:2020cod}
A.~Bose, A.~Sinha, and S.~S. Tiwari, ``{Selection rules for the S-Matrix
  bootstrap},'' \href{http://arxiv.org/abs/2011.07944}{{\ttfamily
  arXiv:2011.07944 [hep-th]}}.

\bibitem{Bose:2020shm}
A.~Bose, P.~Haldar, A.~Sinha, P.~Sinha, and S.~S. Tiwari, ``{Relative entropy
  in scattering and the S-matrix bootstrap},''
  \href{http://dx.doi.org/10.21468/SciPostPhys.9.5.081}{{\em SciPost Phys.}
  {\bfseries 9} (2020) 081}, \href{http://arxiv.org/abs/2006.12213}{{\ttfamily
  arXiv:2006.12213 [hep-th]}}.

\bibitem{Guerrieri:2020bto}
A.~Guerrieri, J.~Penedones, and P.~Vieira, ``{S-matrix Bootstrap for Effective
  Field Theories: Massless Pions},''
  \href{http://arxiv.org/abs/2011.02802}{{\ttfamily arXiv:2011.02802
  [hep-th]}}.

\bibitem{Albert:2022oes}
J.~Albert and L.~Rastelli, ``{Bootstrapping pions at large N},''
  \href{http://dx.doi.org/10.1007/JHEP08(2022)151}{{\em JHEP} {\bfseries 08}
  (2022) 151}, \href{http://arxiv.org/abs/2203.11950}{{\ttfamily
  arXiv:2203.11950 [hep-th]}}.

\bibitem{Fernandez_2023}
C.~Fernandez, A.~Pomarol, F.~Riva, and F.~Sciotti, ``Cornering large-nc qcd
  with positivity bounds,''
  \href{http://dx.doi.org/10.1007/jhep06(2023)094}{{\em Journal of High Energy
  Physics} {\bfseries 2023} no.~6, (June, 2023) }.
  \url{http://dx.doi.org/10.1007/JHEP06(2023)094}.

\bibitem{Albert:2023jtd}
J.~Albert and L.~Rastelli, ``{Bootstrapping Pions at Large $N$. Part II:
  Background Gauge Fields and the Chiral Anomaly},''
  \href{http://arxiv.org/abs/2307.01246}{{\ttfamily arXiv:2307.01246
  [hep-th]}}.

\bibitem{Albert:2023seb}
J.~Albert, J.~Henriksson, L.~Rastelli, and A.~Vichi, ``{Bootstrapping mesons at
  large $N$: Regge trajectory from spin-two maximization},''
  \href{http://arxiv.org/abs/2312.15013}{{\ttfamily arXiv:2312.15013
  [hep-th]}}.

\bibitem{Ma:2023vgc}
T.~Ma, A.~Pomarol, and F.~Sciotti, ``{Bootstrapping the chiral anomaly at large
  N$_{c}$},'' \href{http://dx.doi.org/10.1007/JHEP11(2023)176}{{\em JHEP}
  {\bfseries 11} (2023) 176}, \href{http://arxiv.org/abs/2307.04729}{{\ttfamily
  arXiv:2307.04729 [hep-th]}}.

\bibitem{Bocchia:2026kew}
S.~Bocchia and A.~Vichi, ``{Primal Bootstrap for Pion Scattering at
  Large-Nc},'' \href{http://arxiv.org/abs/2606.14676}{{\ttfamily
  arXiv:2606.14676 [hep-th]}}.

\bibitem{Bocchia:2026lty}
S.~Bocchia and A.~Guerrieri, ``{Large $N$ hadron scattering bootstrap: massive
  pions and glueballs},'' \href{http://arxiv.org/abs/2610.31578}{{\ttfamily
  arXiv:2610.31578 [hep-th]}}.

\bibitem{Guerrieri:2024jkn}
A.~Guerrieri, K.~H{\"a}ring, and N.~Su, ``{From data to the analytic S-matrix:
  A Bootstrap fit of the pion scattering amplitude},''
  \href{http://dx.doi.org/10.21468/SciPostPhys.20.2.034}{{\em SciPost Phys.}
  {\bfseries 20} no.~2, (2026) 034},
  \href{http://arxiv.org/abs/2410.23333}{{\ttfamily arXiv:2410.23333
  [hep-th]}}.

\bibitem{Karateev:2020axc}
D.~Karateev, ``{Two-point functions and bootstrap applications in quantum field
  theories},'' \href{http://dx.doi.org/10.1007/JHEP02(2022)186}{{\em JHEP}
  {\bfseries 02} (2022) 186}, \href{http://arxiv.org/abs/2012.08538}{{\ttfamily
  arXiv:2012.08538 [hep-th]}}.

\bibitem{Chen:2021pgx}
H.~Chen, A.~L. Fitzpatrick, and D.~Karateev, ``{Bootstrapping 2d
  \ensuremath{\phi}$^{4}$ theory with Hamiltonian truncation data},''
  \href{http://dx.doi.org/10.1007/JHEP02(2022)146}{{\em JHEP} {\bfseries 02}
  (2022) 146}, \href{http://arxiv.org/abs/2107.10286}{{\ttfamily
  arXiv:2107.10286 [hep-th]}}.

\bibitem{Correia:2022dyp}
M.~Correia, J.~Penedones, and A.~Vuignier, ``{Injecting the UV into the
  bootstrap: Ising Field Theory},''
  \href{http://dx.doi.org/10.1007/JHEP08(2023)108}{{\em JHEP} {\bfseries 08}
  (2023) 108}, \href{http://arxiv.org/abs/2212.03917}{{\ttfamily
  arXiv:2212.03917 [hep-th]}}.

\bibitem{Cordova:2023wjp}
L.~Cordova, M.~Correia, A.~Georgoudis, and A.~Vuignier, ``{The O(N) monolith
  reloaded: sum rules and Form Factor Bootstrap},''
  \href{http://dx.doi.org/10.1007/JHEP01(2024)093}{{\em JHEP} {\bfseries 01}
  (2024) 093}, \href{http://arxiv.org/abs/2311.03031}{{\ttfamily
  arXiv:2311.03031 [hep-th]}}.

\bibitem{GASSER198765}
J.~Gasser, ``Chiral perturbation theory and effective lagrangians,''
  \href{http://dx.doi.org/https://doi.org/10.1016/0550-3213(87)90307-5}{{\em
  Nuclear Physics B} {\bfseries 279} no.~1, (1987) 65--79}.

\bibitem{Belitsky:2005qn}
A.~V. Belitsky and A.~V. Radyushkin, ``{Unraveling hadron structure with
  generalized parton distributions},''
  \href{http://dx.doi.org/10.1016/j.physrep.2005.06.002}{{\em Phys. Rept.}
  {\bfseries 418} (2005) 1--387},
  \href{http://arxiv.org/abs/hep-ph/0504030}{{\ttfamily arXiv:hep-ph/0504030}}.

\bibitem{Hatta:2018sqd}
Y.~Hatta, A.~Rajan, and K.~Tanaka, ``{Quark and gluon contributions to the QCD
  trace anomaly},'' \href{http://dx.doi.org/10.1007/JHEP12(2018)008}{{\em JHEP}
  {\bfseries 12} (2018) 008}, \href{http://arxiv.org/abs/1810.05116}{{\ttfamily
  arXiv:1810.05116 [hep-ph]}}.

\bibitem{Stegeman:2025tdl}
R.~Stegeman and R.~Zwicky, ``{Gluon gravitational D-form factor: the
  {\ensuremath{\sigma}}-meson as a dilaton confronted with lattice data II},''
  \href{http://dx.doi.org/10.1007/JHEP05(2026)159}{{\em JHEP} {\bfseries 05}
  (2026) 159}, \href{http://arxiv.org/abs/2512.12315}{{\ttfamily
  arXiv:2512.12315 [hep-ph]}}.

\bibitem{Hackett:2023nkr}
D.~C. Hackett, P.~R. Oare, D.~A. Pefkou, and P.~E. Shanahan, ``{Gravitational
  form factors of the pion from lattice QCD},''
  \href{http://dx.doi.org/10.1103/PhysRevD.108.114504}{{\em Phys. Rev. D}
  {\bfseries 108} no.~11, (2023) 114504},
  \href{http://arxiv.org/abs/2307.11707}{{\ttfamily arXiv:2307.11707
  [hep-lat]}}.

\bibitem{narison_2004}
S.~Narison, \href{http://dx.doi.org/10.1017/CBO9780511535000}{{\em QCD as a
  Theory of Hadrons: From Partons to Confinement}}.
\newblock Cambridge Monographs on Particle Physics, Nuclear Physics and
  Cosmology. Cambridge University Press, 2004.

\bibitem{Gasser:1984gg}
J.~Gasser and H.~Leutwyler, ``{Chiral Perturbation Theory: Expansions in the
  Mass of the Strange Quark},''
  \href{http://dx.doi.org/10.1016/0550-3213(85)90492-4}{{\em Nucl. Phys. B}
  {\bfseries 250} (1985) 465--516}.

\bibitem{Pelaez:2004vs}
J.~R. Pelaez and F.~J. Yndurain, ``{The Pion-pion scattering amplitude},''
  \href{http://dx.doi.org/10.1103/PhysRevD.71.074016}{{\em Phys. Rev. D}
  {\bfseries 71} (2005) 074016},
  \href{http://arxiv.org/abs/hep-ph/0411334}{{\ttfamily arXiv:hep-ph/0411334}}.

\bibitem{Colangelo:2001df}
G.~Colangelo, J.~Gasser, and H.~Leutwyler, ``{$\pi \pi$ scattering},''
  \href{http://dx.doi.org/10.1016/S0550-3213(01)00147-X}{{\em Nucl. Phys. B}
  {\bfseries 603} (2001) 125--179},
  \href{http://arxiv.org/abs/hep-ph/0103088}{{\ttfamily arXiv:hep-ph/0103088}}.

\bibitem{Bijnens:1994ie}
J.~Bijnens, G.~Colangelo, and J.~Gasser, ``{K(l4) decays beyond one loop},''
  \href{http://dx.doi.org/10.1016/0550-3213(94)90634-3}{{\em Nucl. Phys. B}
  {\bfseries 427} (1994) 427--454},
  \href{http://arxiv.org/abs/hep-ph/9403390}{{\ttfamily arXiv:hep-ph/9403390}}.

\bibitem{Bijnens:1997vq}
J.~Bijnens, G.~Colangelo, G.~Ecker, J.~Gasser, and M.~E. Sainio, ``{Pion-pion
  scattering at low energy},''
  \href{http://dx.doi.org/10.1016/S0550-3213(97)00621-4}{{\em Nucl. Phys. B}
  {\bfseries 508} (1997) 263--310},
  \href{http://arxiv.org/abs/hep-ph/9707291}{{\ttfamily arXiv:hep-ph/9707291}}.

\bibitem{Bijnens:1998fm}
J.~Bijnens, G.~Colangelo, and P.~Talavera, ``{The Vector and scalar
  form-factors of the pion to two loops},''
  \href{http://dx.doi.org/10.1088/1126-6708/1998/05/014}{{\em JHEP} {\bfseries
  05} (1998) 014}, \href{http://arxiv.org/abs/hep-ph/9805389}{{\ttfamily
  arXiv:hep-ph/9805389}}.

\bibitem{Pelaez:2015qba}
J.~R. Pelaez, ``{From controversy to precision on the sigma meson: a review on
  the status of the non-ordinary $f_0(500)$ resonance},''
  \href{http://dx.doi.org/10.1016/j.physrep.2016.09.001}{{\em Phys. Rept.}
  {\bfseries 658} (2016) 1}, \href{http://arxiv.org/abs/1510.00653}{{\ttfamily
  arXiv:1510.00653 [hep-ph]}}.

\bibitem{TheBook}
B.~Martin, D.~Morgan, and G.~Shaw, {\em {Pion-Pion Interations in particle
  physics}}.
\newblock Academic Press, 1976.

\bibitem{EWERZ201431}
C.~Ewerz, M.~Maniatis, and O.~Nachtmann, ``A model for soft high-energy
  scattering: Tensor pomeron and vector odderon,''
  \href{http://dx.doi.org/https://doi.org/10.1016/j.aop.2013.12.001}{{\em
  Annals of Physics} {\bfseries 342} (2014) 31--77}.

\bibitem{Pelaez:2010er}
J.~R. Pelaez, J.~Nebreda, and G.~Rios, ``{Properties of light resonances from
  unitarized Chiral perturbation theory: Nc behavior and quark mass
  dependence},'' \href{http://dx.doi.org/10.1143/PTPS.186.113}{{\em Prog.
  Theor. Phys. Suppl.} {\bfseries 186} (2010) 113--123},
  \href{http://arxiv.org/abs/1007.3461}{{\ttfamily arXiv:1007.3461 [hep-ph]}}.

\bibitem{Chetyrkin:1996sr}
K.~G. Chetyrkin, ``{Correlator of the quark scalar currents and Gamma(tot) (H
  ---{\ensuremath{>}} hadrons) at O (alpha-s**3) in pQCD},''
  \href{http://dx.doi.org/10.1016/S0370-2693(96)01368-8}{{\em Phys. Lett. B}
  {\bfseries 390} (1997) 309--317},
  \href{http://arxiv.org/abs/hep-ph/9608318}{{\ttfamily arXiv:hep-ph/9608318}}.

\bibitem{Chetyrkin:1996ez}
K.~G. Chetyrkin, ``{Corrections of order alpha-s**3 to R(had) in pQCD with
  light gluinos},'' \href{http://dx.doi.org/10.1016/S0370-2693(96)01478-5}{{\em
  Phys. Lett. B} {\bfseries 391} (1997) 402--412},
  \href{http://arxiv.org/abs/hep-ph/9608480}{{\ttfamily arXiv:hep-ph/9608480}}.

\bibitem{Gorishnii:1990vf}
S.~G. Gorishnii, A.~L. Kataev, and S.~A. Larin, ``{The
  $O(\alpha^{3}_{s})$-corrections to $\sigma_{tot}(e^{+}e^{-}\rightarrow
  hadrons)$ and $\Gamma(\tau^{-} \rightarrow \nu_{\tau} + hadrons)$ in QCD},''
  \href{http://dx.doi.org/10.1016/0370-2693(91)90149-K}{{\em Phys. Lett. B}
  {\bfseries 259} (1991) 144--150}.

\bibitem{Zoller:2012qv}
M.~F. Zoller and K.~G. Chetyrkin, ``{OPE of the energy-momentum tensor
  correlator in massless QCD},''
  \href{http://dx.doi.org/10.1007/JHEP12(2012)119}{{\em JHEP} {\bfseries 12}
  (2012) 119}, \href{http://arxiv.org/abs/1209.1516}{{\ttfamily arXiv:1209.1516
  [hep-ph]}}.

\bibitem{Bagan:1988ay}
E.~Bagan and S.~Narison, ``{Tensor Mesons in {QCD}: Revisited},''
  \href{http://dx.doi.org/10.1016/0370-2693(88)91393-7}{{\em Phys. Lett. B}
  {\bfseries 214} (1988) 451--458}.

\bibitem{He:2021eqn}
Y.~He and M.~Kruczenski, ``{S-matrix bootstrap in 3+1 dimensions:
  regularization and dual convex problem},''
  \href{http://dx.doi.org/10.1007/JHEP08(2021)125}{{\em JHEP} {\bfseries 08}
  (2021) 125}, \href{http://arxiv.org/abs/2103.11484}{{\ttfamily
  arXiv:2103.11484 [hep-th]}}.

\bibitem{cherry2001qcd}
S.~N. Cherry and M.~R. Pennington, ``Qcd finite energy sum rules and the
  isoscalar scalar mesons,'' 2001.

\end{thebibliography}\endgroup

\end{document}